\documentclass[twocolumn]{revtex4-2}

\usepackage[utf8]{inputenc}
\usepackage{amsfonts}
\usepackage{amsmath}
\usepackage{bbm}
\usepackage{amssymb}
\usepackage{graphicx}
\usepackage{xcolor}
\usepackage{siunitx}
\usepackage{float}
\usepackage{mathtools}
\usepackage{fullpage}
\usepackage[hidelinks]{hyperref}
\usepackage{bm}
\usepackage{subdepth}
\usepackage{algorithmicx}
\usepackage{algpseudocode}
\usepackage{algorithm}
\usepackage[mathlines, modulo]{lineno}

\newcommand*{\T}[1]{\mathcal{T}^{#1}}
\newcommand*{\sT}[1]{\mathcal{S}^{#1}}

\DeclareMathOperator{\E}{\mathbbm{E}}

\DeclarePairedDelimiter{\set}{\{}{\}}
\DeclarePairedDelimiter{\norm}{\lVert}{\rVert}
\DeclarePairedDelimiterX{\inner}[2]{\langle}{\rangle}{{#1},{#2}}

\newcommand*{\ind}[1]{\mathbbm{1}_{#1}}
\newcommand*{\bmin}{\wedge}                 
\newcommand*{\comp}{\mathsf{c}}  

\begin{document}

\title{Inexact iterative numerical linear algebra for neural network-based spectral estimation and rare-event prediction}

\date{\today}

\author{John Strahan}
\affiliation{Department of Chemistry and James Franck Institute, University of Chicago, Chicago, Illinois 60637, United States}
\author{Spencer C. Guo}
\affiliation{Department of Chemistry and James Franck Institute, University of Chicago, Chicago, Illinois 60637, United States}
\author{Chatipat Lorpaiboon}
\affiliation{Department of Chemistry and James Franck Institute, University of Chicago, Chicago, Illinois 60637, United States}
\author{Aaron R. Dinner}
\email{dinner@uchicago.edu}
\affiliation{Department of Chemistry and James Franck Institute, University of Chicago, Chicago, Illinois 60637, United States}
\author{Jonathan Weare}
\email{weare@nyu.edu}
\affiliation{Courant Institute of Mathematical Sciences, New York University, New York, New York 10012, United States}

\begin{abstract}
    Understanding dynamics in complex systems is challenging because there are many degrees of freedom, and those that are most important for describing events of interest are often not obvious.  The leading eigenfunctions of the transition operator
    are useful for visualization, and they can provide an efficient basis for computing statistics such as the likelihood and average time of events (predictions).  Here we  develop inexact iterative linear algebra methods for computing these eigenfunctions (spectral estimation) and making predictions from a data set of short trajectories sampled at finite intervals.  We demonstrate the methods on a low-dimensional model that facilitates visualization and a high-dimensional model of a biomolecular system.  Implications for the prediction problem in reinforcement learning are discussed.
\end{abstract}

\maketitle

\section{Introduction}

Modern observational, experimental, and computational approaches often yield high-dimensional time series data (trajectories) for complex systems.  In principle, these trajectories contain rich information about dynamics and, in particular, the infrequent events that are often most consequential.
In practice, however, high-dimensional trajectory data are often difficult to parse for useful insight. The need for more efficient statistical analysis tools for trajectory data is critical, especially when the goal is to understand rare-events that may not be well represented in the data.

We consider dynamics that can be treated as Markov processes.  A common starting point for statistical analyses of Markov processes is the transition operator, which describes the evolution of function expectations.  The eigenfunctions of the transition operator characterize the most slowly decorrelating features (modes) of the system \cite{noe_variational_2013, nuske_variational_2014, klus2018data, webber2021error,lorpaiboon_integrated_2020}. 
These can be used for dimensionality reduction to obtain a qualitative understanding of the dynamics 
 \cite{mcgibbon_identification_2017,busto2021structural}, or they can be used as the starting point for further computations \cite{perez-hernandez_identification_2013, schwantes_improvements_2013,strahan_long-time-scale_2021}.  Similarly, prediction functions, which provide information about the likelihood and timing of future events as a function of the current state, are defined through linear equations of the transition operator \cite{thiede_galerkin_2019,strahan_long-time-scale_2021}.

A straightforward numerical approach to obtaining these functions is to convert the transition operator to a matrix by projecting onto a finite basis for Galerkin approximation \cite{swope_describing_2004, noe_constructing_2009, bowman_introduction_2014, noe_variational_2013, nuske_variational_2014, thiede_galerkin_2019, strahan_long-time-scale_2021, finkel_learning_2021}.
The performance of such a linear approximation depends on the choice of basis \cite{thiede_galerkin_2019, strahan_long-time-scale_2021,finkel_learning_2021}, and previous work often resorts to a set of indicator functions on a partition of the state space (resulting in a Markov state model or MSM \cite{bowman_introduction_2014}) for lack of a better choice.
While Galerkin approximation has yielded many insights \cite{antoszewski_kinetics_2021,guo_dynamics_2022}, the limited expressivity of the basis expansion has stimulated interest in (nonlinear) alternatives.


In particular, artificial neural networks can be harnessed to learn eigenfunctions of the transition operator and prediction functions from data \cite{andrew_deep_2013, mardt_vampnets_2018, wehmeyer_time-lagged_2018, lusch_deep_2018, chen_nonlinear_2019, lorpaiboon_integrated_2020, glielmo_unsupervised_2021, strahan2023predicting, li_semigroup_2022}.  
However, existing approaches based on neural networks suffer from various drawbacks.
As discussed in Ref.~\onlinecite{lorpaiboon_integrated_2020}, their performance can often be very sensitive to hyperparameters, requiring extensive tuning and varying with random initialization.  Many use loss functions that are estimated against the stationary distribution \cite{khoo_solving_2018, li_computing_2019, roux_string_2021, li_semigroup_2022, roux_transition_2022, rotskoff_active_2022}, so that metastable states contribute most heavily, which negatively impacts performance \cite{rotskoff_active_2022,strahan2023predicting}.  Assumptions about the dynamics (e.g., microscopic reversibility) limit applicability.
In Ref.~\onlinecite{strahan2023predicting} we introduced an approach that overcomes the issues above, but it uses multiple trajectories from each initial condition; this limits the approach to analysis of simulations and moreover requires specially prepared data sets.

The need to compute prediction functions from observed trajectory data also arises in reinforcement learning.
There the goal is to optimize an expected future reward (the prediction function) over a policy (a Markov process). For a fixed Markov process, the prediction problem in reinforcement learning is often solved by temporal difference (TD) methods, which allow the use of arbitrary ensembles of trajectories without knowledge of the details of the underlying dynamics \cite{sutton_reinforcement_2018}.  TD methods have a close relationship with an inexact form of power iteration, which, as we describe, can perform poorly on rare-event related problems.

Motivated by this relationship, as well as by an inexact power iteration scheme previously proposed for approximating the stationary probability distribution of a Markov process using trajectory data~ \cite{wen_batch_2020}, we propose a computational framework for spectral estimation and rare-event prediction based on inexact iterative numerical linear algebra. 
Our framework includes an inexact Richardson iteration for the prediction problem, as well as an extension to inexact subspace iteration for the prediction and spectral estimation problems.
The theoretical properties of exact subspace iteration suggest that eigenfunctions outside the span of the approximation will contribute significantly to the error of our inexact iterative schemes \cite{GoluVanl96}. Consistent with this prediction, we demonstrate that learning additional eigenvalues and eigenfunctions simultaneously through inexact subspace iteration accelerates convergence dramatically relative to inexact Richardson and power iteration in the context of rare events. 
While we assume the dynamics can be modeled by a Markov process, we do not require knowledge of their form or a specific underlying model.
The method shares a number of further advantages with the approach discussed in Ref.\ \onlinecite{strahan2023predicting} without the need for multiple trajectories from each initial condition in the data set.  This opens the door to treating a wide range of observational, experimental, and computational data sets.  

The remainder of the paper is organized as follows. In Section~\ref{sec:setup}, we describe the quantities that we seek to compute in terms of linear operators.  In Sections~\ref{sec:inexactPower} and \ref{sec:ieSI}, we introduce an inexact subspace iteration algorithm that we use to solve for these quantities. Section~\ref{sec:softplus} illustrates how the loss function can be tailored to the known properties of the desired quantity.
Section~\ref{sec:testproblems} summarizes the two test systems that we use to illustrate our methods: a two-dimensional potential, for which we can compute accurate reference solutions, and a molecular example that is high-dimensional but still sufficiently tractable that statistics for comparison can be computed from long trajectories.  In Section~\ref{sec:eig}, we explain the details of the invariant subspace iteration and then demonstrate its application to our two examples.  Lastly, Section~\ref{sec:forecast} details how the subspace iteration can be modified to compute prediction functions and compares the effect of different loss functions, as well as the convergence properties of power iteration and subspace iteration. 
We conclude with implications for reinforcement learning.



\section{Spectral estimation and the prediction problem}\label{sec:setup}
We have two primary applications in mind in this article.  First, we would like to estimate the \emph{dominant eigenfunctions and eigenvalues} of the transition operator $\T{t}$ for a Markov process $X^t\in \mathbb{R}^d$, defined as
\begin{equation}\label{eq:to}
    \T{t} f(x) = \E_x\left[ f(X^t) \right],
\end{equation}
where $f$ is an arbitrary real-valued function and the subscript $x$ indicates the initial condition $X^0=x$. The transition operator encodes how expectations of functions evolve in time.   The right eigenfunctions of $\T{t}$ with the largest eigenvalues characterize the most slowly decorrelating features (modes) of the Markov process \cite{noe_variational_2013, nuske_variational_2014,webber2021error,lorpaiboon_integrated_2020}.

Our second application is to compute \emph{prediction functions} of the general form 
\begin{equation}\label{eq:forecast}
    u(x) = \E_x\left[ \Psi(X^T) + \sum_{t=0}^{T-1} \Gamma(X^t)\right],
\end{equation}
where $T$ is the first time $X^t\notin D$ from some domain $D$, and $\Psi$ and $\Gamma$ are functions associated with the escape event (rewards in reinforcement learning).  Prototypical examples of prediction functions that appear in our numerical results are the mean first passage time (MFPT) from each $x$:
\begin{equation}\label{eq:mfpt}
m(x) = \mathbbm{E}_x\left[ T  \right]
\end{equation}
and the committor:
\begin{equation}\label{eq:committor}
q(x) = \mathbbm{E}_x\left[ \mathbbm{1}_B(X^T)  \right]=\mathbbm{P}_x\left[ X^T \in B \right],
\end{equation}
where $A$ and $B$ are disjoint sets (``reactant'' and ``product'' states) and $D= (A\cup B)^\comp$.
The MFPT is important for estimating rates of transitions between regions of state space, while the committor can serve as a reaction coordinate \cite{du_transition_1998, Ma2005nn, roux_transition_2022, krivov_reaction_2013} and as a key ingredient in transition path theory statistics~ \cite{strahan2023predicting, e_transition-path_2010, vanden2006transition}.
For any $\tau>0$, the prediction function $u(x)$ satisfies the linear equation
\begin{equation}\label{eq:FK}
    \left( \mathcal{I} - \sT{\tau}\right) u(x)  
    = \E_x \left[ \sum_{t=0}^{(\tau\bmin T)-1} \Gamma(X^t)\right]
\end{equation}
for $x\in D$, with boundary condition
\begin{equation}
u(x) = \Psi(x)
\end{equation}
for $x\notin D$.
In \eqref{eq:FK}, $\mathcal{I}$ is the identity operator and 
\begin{equation}
    \mathcal{S}^\tau f(x) = \E_x\left[ f(X^{\tau\wedge T}) \right]
\end{equation}
is the ``stopped'' transition operator \cite{strahan_long-time-scale_2021}. We use the notation $\tau\wedge T = \min\{\tau, T\}$.  
The parameter $\tau$ is known as the lag time. While
it is an integer corresponding to a number of integration steps, in
our numerical examples, we often express it in terms of equivalent
time units.


Our specific goal is to solve both the eigenproblem and the prediction problem for $X^t$ in high dimensions and without direct access to a model governing its evolution.  Instead, we have access to trajectories of $X^t$ of a fixed, finite duration $\tau$.
A natural and generally applicable approach to finding an approximate solution to the prediction problem is to attempt to minimize the residual of \eqref{eq:FK} over parameters $\theta$ of a neural network $u_\theta(x)$ representing $u(x)$.  
For example, we recently suggested an algorithm that minimizes the residual norm 
\begin{equation}\label{eq:Res}
\frac{1}{2}\norm[\Big]{\left( \mathcal{I} - \sT{\tau}\right) u_\theta - r }_\mu^2,
\end{equation}
where $r(x)$ is the right-hand side of \eqref{eq:FK} and $\mu$ is an arbitrary distribution of initial conditions $X^0$ (boundary conditions were enforced by an additional term) \cite{strahan2023predicting}. 
The norm $\lVert\cdot\rVert_\mu$ is defined by $\lVert f\rVert_\mu^2=\langle f,f\rangle_\mu$, where $\langle f,g\rangle_\mu=\int f(x)g(x)\mu(dx)$ for arbitrary functions $f$ and $g$.
The gradient of the residual norm in \eqref{eq:Res} with respect to neural-network parameters $\theta$ can be written
\begin{widetext}
\begin{equation}\label{eq:gradres}
 \inner[\big]{\left( \mathcal{I} - \mathcal{S}^\tau\right) u_\theta - r}{\nabla_\theta u_\theta }_\mu - 
\inner[\big]{\left( \mathcal{I} - \mathcal{S}^\tau\right) u_\theta - r}{\mathcal{S}^\tau \nabla_\theta u_\theta}_\mu
\end{equation}
Given a data set of initial conditions ${\{X^0_j\}}_{j=1}^n$ drawn from $\mu$ and a single sample trajectory ${\{X^t_j\}}_{t=0}^\tau$ of $X^t$ for each $X^0_j$, the first term in the gradient \eqref{eq:gradres} can be approximated without bias as
\begin{equation}\label{eq:semigrad}
\inner[\big]{\left( \mathcal{I} - \mathcal{S}^\tau\right) u_\theta - r} { \nabla_\theta u_\theta }_\mu \approx \frac{1}{n} \sum_{j=1}^n \left(u_\theta(X^0_j) - u_\theta(X^{\tau\bmin T_j}_j) - \sum_{t=0}^{(\tau\bmin T_j)-1} \Gamma(X^t_j)\right)\nabla_\theta u_\theta(X_j^0)
\end{equation}
\end{widetext}
where $T_j$ is the first time  $X^t_j\notin D$.

In contrast, unbiased estimation of the second term in \eqref{eq:gradres} requires access to two independent trajectories of $X^t$ for each sample initial condition since it is quadratic in $\mathcal{S}^\tau$~ \cite{strahan2023predicting,sutton_reinforcement_2018}. In the reinforcement learning community, TD methods were developed for the purpose of minimizing a loss of a very similar form to \eqref{eq:Res}, and they perform a ``semigradient'' descent by following only the first term in \eqref{eq:gradres} \cite{sutton_reinforcement_2018}.
As in the semigradient approximation, the algorithms proposed in this paper only require access to inner products of the form ${\inner{ f }{ \mathcal{A} g}}_\mu$ for an operator $\mathcal{A}$ related to $\T{\tau}$ or $\sT{\tau}$, and they avoid terms that are non-linear in $\mathcal{A}$.  As we explain, such inner products can be estimated using trajectory data.
However, rather than attempting to minimize the residual directly by an approximate gradient descent, we view the eigenproblem and prediction problems through the lens of iterative numerical linear algebra schemes. 

\section{An inexact power iteration}\label{sec:inexactPower}

To motivate the iterative numerical linear algebra point of view, observe that the first term in \eqref{eq:gradres} is the exact gradient with respect to $\theta'$ of the loss\begin{equation}\label{eq:powerloss}
\frac{1}{2}\norm[\big]{ u_{\theta'} - \mathcal{S}^\tau u_{\theta} - r }_\mu^2,
\end{equation}
evaluated at $\theta'=\theta$.  
The result of many steps of gradient descent (later, ``inner iterations'') on this loss with $\theta$ held fixed can then be viewed as 
an inexact Richardson iteration for \eqref{eq:FK}, resulting in a sequence
\begin{equation}\label{eq:ierich}
u_{\theta^{s+1}} \approx \mathcal{S}^\tau u_{\theta^s} + r,
\end{equation}
where, for each $s$,  $u_{\theta^s}$ is a  parametrized neural-network approximation of the solution to \eqref{eq:FK}.
To unify our discussion of the prediction and spectral estimation problems, it is helpful to observe that  \eqref{eq:ierich} can be recast as an equivalent inexact power iteration:
\begin{equation}\label{eq:iepow}
    \bar u_{\theta^{s+1}} \approx \mathcal{A}\bar u_{\theta^s}
\end{equation}
where
\begin{equation}
\label{eq:A1}
\bar u_\theta = \left(\begin{array}{c}
1\\
u_\theta
\end{array}\right)
\quad\text{and}
\quad
\mathcal{A} = \left[
\begin{array}{cc}
1 & 0\\
r & \mathcal{S}^\tau
\end{array}
\right].
\end{equation}
Note that $(1, u)^\top$ is the dominant eigenfunction of $\mathcal{A}$ and has eigenvalue equal to 1.

Ref.~\onlinecite{wen_batch_2020} introduced an inexact power iteration to compute the stationary probability measure of $\mathcal{T}^\tau$, i.e., its dominant left eigenfunction.  As those authors note, an inexact power update such as \eqref{eq:iepow} can  be enforced using a variety of loss functions.
In our setting, the $L^2_\mu$ norm in \eqref{eq:powerloss} can be replaced by any other measure of the difference between $u_{\theta'}$ and $\mathcal{S}^\tau u_{\theta} + r$, as long as  an unbiased estimator of its gradient with respect to $\theta'$ is available.
This flexibility is discussed in more detail in Section \ref{sec:softplus}, and we exploit it in applications presented later in this article.
For now, we focus instead on another important implication of this viewpoint: the flexibility in the form of the iteration itself. 

We will see that when the spectral gap of $\mathcal{A}$, the difference between its largest and second largest eigenvalues, is small, the inexact power iteration (or the equivalent Richardson iteration) described above fails to reach an accurate solution.
The largest eigenvalue of $\mathcal{A}$ in \eqref{eq:A1} is 1 and its next largest eigenvalue is the dominant eigenvalue of $\mathcal{S}^\tau$.
For rare-event problems the difference between these two eigenvalues is expected to be very small. Indeed, when $X^0$ is drawn from the quasi-stationary distribution of $X^t$ in $D$, the logarithm of the largest eigenvalue of $\mathcal{S}^\tau$ is exactly $-\tau/\E[T]$ \cite{collett_quasi-stationary_2012}. Thus, when the mean exit time from $D$ is large, we can expect the spectral gap of $\mathcal{A}$ in \eqref{eq:A1} to be very small. 
In this case, classical convergence results tell us that exact power iteration converges slowly, with the largest contributions to the error coming from the eigenfunctions of $\mathcal{S}^\tau$ with largest magnitude eigenvalues~ \cite{GoluVanl96}.  Iterative schemes that approximate multiple dominant eigenfunctions simultaneously, such as subspace iteration and Krylov methods, can converge much more rapidly \cite{GoluVanl96}.
In the next section, we introduce an inexact form of subspace iteration to address this shortcoming.  Beyond dramatically improving performance on the prediction problem for rare-events when applied with $\mathcal{A}$ in \eqref{eq:A1}, it can also be applied with $\mathcal{A}=\mathcal{T}^\tau$ to compute multiple dominant eigenfunctions and values of  $\mathcal{T}^\tau$ itself.

\section{An inexact subspace iteration}\label{sec:ieSI}


Our inexact subspace iteration for the $k$ dominant eigenfunctions/values of $\mathcal{A}$ proceeds as follows.  Let $\set{\varphi^a_{\theta^s}}_{a=1}^k$ be a sequence of $k$ basis functions parametrized by $\theta^s$ (these can be scalar or vector valued functions depending on the form of $\mathcal{A}$). We can represent this basis as the vector valued function \begin{equation}\label{eq:U}
    U_{\theta^s} = \left(\varphi^1_{\theta^s},\varphi^2_{\theta^s},\dots,\varphi^k_{\theta^s}\right).
\end{equation}  
Then, we can obtain a new set of $k$ basis functions by approximately applying $\mathcal{A}$ to each of the components of $U_{\theta^s}$:
\begin{equation}\label{eq:ieSI}
U_{\theta^{s+1}}K^{s+1} \approx \mathcal{A}\, U_{\theta^s}
\end{equation}
where $K^{s+1}$ is an invertible, upper-triangular $k\times k$ matrix that does not change the span of the components of $U_{\theta^{s+1}}$ but is included to facilitate training.
One way the approximate application of $\mathcal{A}$ can be accomplished is by minimizing 
\begin{equation}\label{eq:general_Loss}
\frac{1}{2}\sum_{a=1}^k \norm[\Bigg]{ \sum_{b=1}^k \varphi^b_{\theta}K_{ba} - \mathcal{A}\, \varphi^a_{\theta^s} }^2_\mu
\end{equation}
over $\theta$ and $K$
with $\theta^s$ held fixed.

The eigenvalues and eigenfunctions of $\mathcal{A}$ are then approximated by solving the finite-dimensional generalized eigenproblem
\begin{equation}
    C^\tau W = C^0 W \Lambda
\end{equation}
where
\begin{align}
    C^\tau_{ab} &= {\langle \varphi_{\theta^s}^a, \mathcal{A} \varphi_{\theta^s}^b\rangle}_\mu \label{eq:ct}\\
    C^0_{ab} & =  {\langle \varphi_{\theta^s}^a, \varphi_{\theta^s}^b\rangle}_\mu \label{eq:c0},
\end{align}
 each inner product is estimated using trajectory data, and 
$W$ and $\Lambda$ are $k\times k$ matrices. The matrix $\Lambda$ is diagonal, and the $a$-th eigenvalue $\lambda_a$ of $\mathcal{A}$ is  approximated by $\Lambda_{aa}$; the corresponding eigenfunction  $v_a$ is 
approximated by $\sum_{b=1}^k W_{ab}\, \varphi_{\theta^s}^b$.

Even when sampling is not required to estimate the matrices in \eqref{eq:ct} and \eqref{eq:c0}, the numerical rank of $C^\tau$ becomes very small as the eigenfunctions become increasingly aligned with the single dominant eigenfunction.  To overcome this issue, we apply an orthogonalization step between iterations (or every few iterations). 
Just as the matrices $C^0$ and $C^\tau$ can be estimated using trajectory data, the orthogonalization procedure can also be implemented approximately using data. 

Finally, in our experiments we find it advantageous to damp large parameter fluctuations during training by mixing the operator $\mathcal{A}$ with a multiple of the identity, i.e., we perform our inexact subspace iteration on the operator $(1-\alpha_s) \mathcal{I} + \alpha_s \mathcal{A}$ in place of $\mathcal{A}$ itself.  This new operator has the same eigenfunctions as $\mathcal{A}$.  In our experiments, decreasing the parameter $\alpha_s$ as the number of iterations increases results in better generalization properties of the final solution and helps ensure convergence of the iteration.
For our numerical experiments we use 
\begin{equation}
    \alpha_s =
    \begin{cases}
    1 \qquad & s < \sigma\\
    1 / \sqrt{s +1 - \sigma} \qquad & s\geq \sigma
    \end{cases}
\end{equation}
where $\sigma$ is a user chosen hyperparameter that sets the number of iterations performed before damping begins.

The details, including estimators for all required inner products, in the case of the eigenproblem ($\mathcal{A}=\mathcal{T}^\tau$) are given in Section~\ref{sec:eig} and Algorithm~\ref{alg:eig}. For the prediction problem with $\mathcal{A}$ as in \eqref{eq:A1}, they are given in Section~\ref{sec:forecast} and Algorithm~\ref{alg:forecast}.

\section{Alternative loss functions}\label{sec:softplus}

As mentioned above, the inexact application of the operator $\mathcal{A}$ can be accomplished by minimizing loss functions other than \eqref{eq:general_Loss}.  The key requirement for a loss function in the present study is that $\mathcal{A}$ appears in its gradient only through terms of the form $\langle f, \mathcal{A} g\rangle_\mu$ for some functions $f$ and $g$, so that the gradient can be estimated using trajectory data.
As a result, we have flexibility in the choice of loss and, in turn, the representation of $u$.  In particular, we consider the representation $u_\theta = \zeta(z_\theta)$, where $\zeta$ is an increasing function, and $z_\theta$ is a function parameterized by a neural network. An advantage of doing so is that the function $\zeta$ can restrict the output values of $u_\theta$ to some range.  For example, when computing a probability such as the committor, a natural choice is the sigmoid function $\zeta(x) = (1+e^{-x})^{-1}$. 

Our goal is to train a sequence of parameter values so that $u_{\theta^s}$ approximately follows a subspace iteration, i.e., so that $\zeta(z_{\theta^{s+1}}) \approx \mathcal{A} u_{\theta^s}$.
To this end, we minimize with respect to $\theta$ the loss function
\begin{equation}\label{eq:genloss}
\E_{X^0\sim\mu}\left[ V(z_\theta)
- z_\theta \mathcal{A}u_{\theta^s}\right],
\end{equation}
where $V$ is an antiderivative of $\zeta$, and $\theta^s$ is fixed. The subscript $X^0\sim \mu$ in this expression indicates that $X^0$ is drawn from $\mu$.
Note that, as desired, $\mathcal{A}$ appears in the gradient of \eqref{eq:genloss} only in an inner product of the required form, and an approximate minimizer, $\theta^{s+1}$, of this loss satisfies $\zeta(z_{\theta^{s+1}}) \approx \mathcal{A} u_{\theta^s}$.  This general form of loss function is adapted from variational expressions for the divergence of two probability measures~ \cite{Nguyen2010div,wen_batch_2020}.

The $L^2_\mu$ loss in \eqref{eq:general_Loss}, which we use in several of our numerical experiments, corresponds to the choice $\zeta(x) = x$ and $V(x)= x^2/2$.  The choice of $\zeta(x) = (1+e^{-x})^{-1}$ mentioned above corresponds to $V(x) = \log(1+e^x)$; we refer to the loss in \eqref{eq:genloss} with this choice of $V$ as the ``softplus'' loss. 
{
We note that in the context of committor function estimation, in the limit of infinite lag time the ``softplus'' loss corresponds directly to the maximum log-likelihood loss (for independent Bernoulli random variables) that defines the logistic regression estimate of the probability of reaching $B$ before $A$. Logistic regression has previously been used in conjunction with data sets of trajectories integrated all the way until reaching $A$ or $B$ to estimate the committor function \cite{peters_obtaining_2006, peters2007extensions, jung_artificial_2019, chattopadhyay2020analog,jung_machine-guided_2023,  miloshevich_probabilistic_2023}.
}

\section{Test problems}\label{sec:testproblems}

We illustrate our methods with two well-characterized systems that exemplify features of molecular transitions.  In this section, we provide key details of these systems.

\begin{figure}[bt]
    \centering
    \includegraphics[width=0.8\linewidth]{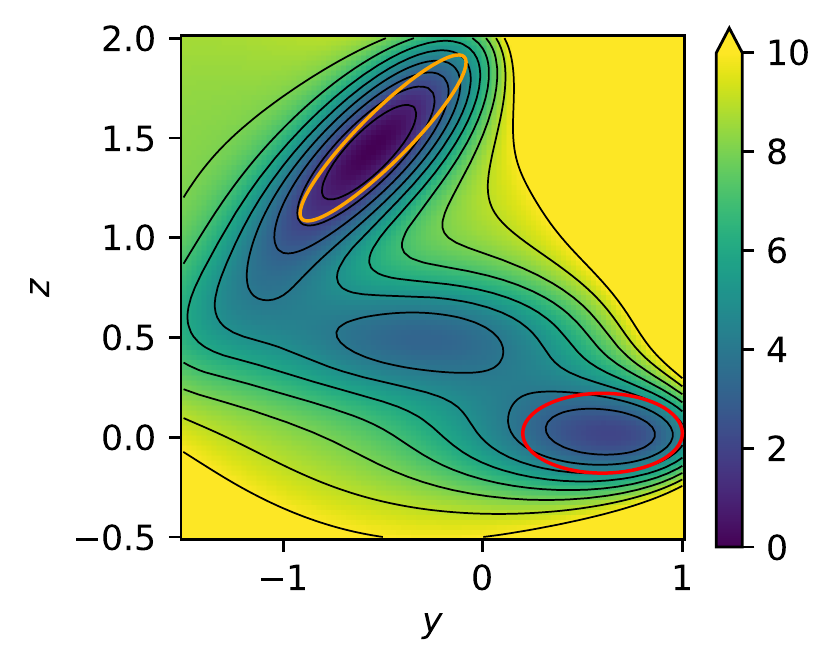}
    \caption{M\"uller-Brown potential energy surface.  The orange and red ovals indicate the locations of states $A$ and $B$ respectively when computing predictions. Contour lines are drawn every 1 $\beta^{-1}$.}
    \label{fig:MB}
\end{figure}

\subsection{M\"uller-Brown potential}\label{sec:mb_numerics}

The first system is defined by the M\"uller-Brown potential \cite{muller_location_1979} (Figure~\ref{fig:MB}): 
\begin{multline}
\label{eq:MB}
V_{\rm MB}(y, z)=\frac{1}{20}\sum_{i=1}^4C_i \exp[a_i(y-y_i)^2\\+b_i(y-y_i)(z-z_i)+c_i(z-z_i)^2].
\end{multline}
The two-dimensional nature of this model facilitates visualization. The presence of multiple minima and saddlepoints that are connected by a path that does not align with the coordinate axes makes the system challenging for both sampling and analysis methods.  
In Sections \ref{sec:mbeigenfunctions} and \ref{sec:mbcommittor}, we use $C_i=\{-200,-100,-170,15\}$, $a_i=\{-1,-1,-6.5,0.7\}$,
$b_i=\{0,0,11,0.6\}$, 
$c_i=\{-10,-10,-6.5,0.7\}$, 
$y_i=\{1,-0.27,-0.5,-1\}$, 
$z_i=\{0,0.5,1.5,1\}$.
In Section~\ref{sec:Mod_MB}, we tune the parameters to make transitions between minima rarer; there, the parameters are $C_i=\{-250,-150,-170,15\}$, $a_i=\{-1,-3,-6.5,0.7\}$,
$b_i=\{0,0,11,0.6\}$, 
$c_i=\{-10,-30,-6.5,0.7\}$, 
$y_i=\{1,-0.29,-0.5,-1\}$, 
$z_i=\{0,0.5,1.5,1\}$.
For prediction, we analyze transitions between the upper left minimum ($A$) and the lower right minimum ($B$) in Figure \ref{fig:MB}; these states are defined as
\begin{widetext}
\begin{equation}
\label{eqn:MBstates}
\begin{aligned}
&A=\set{ y, z : 6.5(y+0.5)^2-11(y+0.5)(z-1.5)+6.5(z-1.5)^2<0.3}\\
&B=\set{y, z:(y-0.6)^2+0.5(z-0.02)^2<0.2}.
\end{aligned}
\end{equation}

To generate a data set, we randomly draw 50,000 initial conditions $X^0_j$ uniformly from the region 
\begin{equation}
\label{eq:initregion}
\Omega=\set{y, z :-1.5<y<1.0,\ -0.5<z<2.5,\ V_{\rm{MB}}(y,z)<12}
\end{equation}    
\end{widetext}
and, from each of these initial conditions,  generate one trajectory according to the discretized overdamped Langevin dynamics
\begin{equation}
    X^{t}_j = X^{t-1}_j - {\delta_t}\,\nabla V_\textrm{MB}(X^{t-1}_j) + \sqrt{{\delta_t}\,2  \beta^{-1}}\, \xi^{t}_j 
\end{equation}
where $1\leq t\leq \tau$, the $\xi^t_j$ are independent standard Gaussian random variables, and the timestep is ${\delta_t}=0.001$.
We use an inverse temperature of $\beta = 2$, and we vary $\tau$ as indicated below (note that $\tau$ is an integer number of steps, but we present our results for the M\"uller-Brown model in terms of the dimensionless scale used for $\delta_t$ immediately above).  { For each test, we use independent random numbers and redraw the initial conditions because it is faster to generate the trajectories than to read them from disk in this case.  All error bars are computed from the empirical standard deviation over 10 replicate data sets.}

To validate our results, we compare the neural-network results against grid-based references, computed as described in 
{
Section S4 of the Supplementary Material of Ref.\ \onlinecite{thiede_galerkin_2019} and the Appendix of Ref.\ \onlinecite{lorpaiboon_augmented_2022} (our notation here follows the former more closely).
The essential idea is that the terms in the infinitesimal generator of the transition operator can be approximated on a grid to second order in the spacing by expanding both the probability for transitions between sites and a test function.
To this end, we define the transition matrix for neighboring sites  $x = (y,z)$  and $x'=(y\pm \delta_x,z)$ or $(y,z\pm \delta_x)$ on a grid with spacings $\delta_x$:
\begin{equation}
P(x'|x)=\frac{1}{1+e^{\beta[V(x')-V(x)]}}
\end{equation}
(this definition differs from that in Ref.\ \onlinecite{thiede_galerkin_2019} by a factor of 1/4, and we scale $P-I$, where $I$ is the identity matrix, accordingly to set the time units below).
The diagonal entry $P(x|x)$ is fixed to make the transition matrix row-stochastic.  We use $\delta_x=0.0125$; the grid is a rectangle with $y \in [-1.5,1.0]$, and $z \in [-0.5,2.0]$.
The reference invariant subspaces are computed by diagonalizing the matrix $2(P-I)/\beta \delta_x^2$ with a sparse eigensolver; we use scipy.sparse.linalg.
We obtain the reference committor $\hat{q}_+$ for grid sites in $(A\cup B)^\comp$ by solving
\begin{multline}
\mathrm{diag}(\hat{\mathbbm{1}}_{(A\cup B)^\comp})(P-I)\hat{q}_+\\=-\mathrm{diag}(\hat{\mathbbm{1}}_{(A\cup B)^\comp})(P-I)\hat{\mathbbm{1}}_B
\end{multline}
and for grid sites in $A\cup B$ by setting $\hat{q}_+=\hat{\mathbbm{1}}_B$.
Here, $\hat{q}_+$ and $\hat{\mathbbm{1}}_B$ are vectors corresponding to the functions evaluated at the grid sites.
}



\subsection{\texorpdfstring{AIB\textsubscript{9}}{AIB9} helix-to-helix transition}
\label{sec:aib_numerics}

\begin{figure}[tb]
    \centering
    \includegraphics[width=\linewidth]{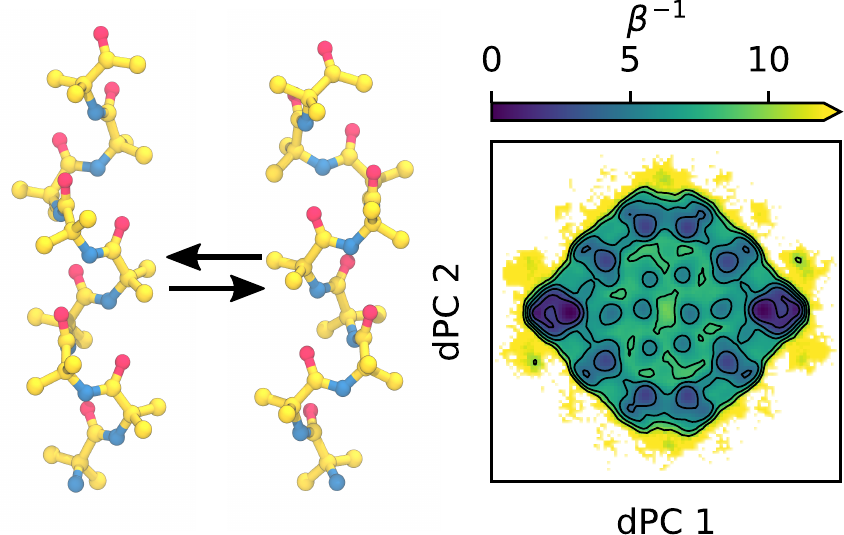}
    \caption{Helix-to-helix transition of AIB\textsubscript{9}.  (left) Left- and right-handed helices, which we use as states $A$ and $B$, respectively, when computing predictions.  Carbon, nitrogen, and oxygen atoms are shown in yellow, blue, and red, respectively; hydrogen atoms are omitted.
    (right) Potential of mean force constructed from the histogram of value pairs of the first two dihedral angle principal components; data are from the 20 trajectories of 5 $\mu$s that we use to construct reference statistics (see text). The left-handed helix corresponds to the left-most basin, and the right-handed helix corresponds to the right-most basin. Contour lines are drawn every 2 $\beta^{-1}$ corresponding to a temperature of 300 K.}
    \label{fig:aib9}
\end{figure}

The second system is a peptide of nine $\alpha$-aminoisobutyric acids (AIB\textsubscript{9}; Figure~\ref{fig:aib9}).  Because AIB is achiral around its $\alpha$-carbon atom, AIB\textsubscript{9} can form left- and right-handed helices with equal probabilities, and we study the transitions between these two states.  This transition was previously  studied using MSMs and long unbiased molecular dynamics simulations~ \cite{buchenberg_hierarchical_2015, perez_meld-path_2018}. AIB\textsubscript{9} poses a stringent test due to the presence of many metastable intermediate states.

The states are defined in terms of the internal $\phi$ and $\psi$ dihedral angles.  We classify an amino acid as being in the ``left'' state if its dihedral angle values are within a circle of radius $25^\circ$ centered at $(41^\circ, 43^\circ)$, that is
\begin{equation*}
    (\phi - 41^\circ)^2 + (\psi - 43^\circ)^2 \le (25^\circ)^2.
\end{equation*}
Amino acids are classified as being in the ``right'' state using the same radius, but centered instead at $(-41^\circ, -43^\circ)$.
States $A$ and $B$ are defined by the amino acids at sequence positions 3--7 being all left or all right, respectively.
We do not use the two residues on each end of AIB\textsubscript{9} in defining the states as these are typically more flexible \cite{perez_meld-path_2018}.
The states can be resolved by projecting onto dihedral angle principal components (dPCs; Figure~\ref{fig:aib9}, right) as described previously \cite{sittel_principal_2017}.

Following a procedure similar to that described in Ref.~ \citenum{perez_meld-path_2018}, we generate a data set of short trajectories.
{ From each of the 691 starting configurations in Ref.~ \citenum{perez_meld-path_2018}, we simulate 10 trajectories of duration 20~ns with initial velocities drawn randomly from a Maxwell-Boltzmann distribution at a temperature of 300~K.  The short trajectory data set thus contains 6,910 trajectories, corresponding to a total sampling time of 138.2 $\mu$s.}
We use a timestep of 4~fs together with a hydrogen mass repartitioning scheme~ \cite{hopkins_long-time-step_2015}, and configurations are saved every 40~ps.
We employ the AIB parameters from Forcefield\_NCAA \cite{khoury_forcefield_ncaa_2014} and the GBNeck2 implicit-solvent model~ \cite{nguyen_improved_2013}.
Simulations are performed with the Langevin integrator in OpenMM 7.7.0 \cite{eastman_openmm_2017} using a friction coefficient of \SI{1}{\per\pico\second}.
To generate a reference for comparison to our results, we randomly select 20 configurations from the data set above and, from each, run a simulation of 5 $\mu$s with the same simulation parameters.
For all following tests on AIB\textsubscript{9}, batches consist of pairs of frames separated by $\tau$ drawn randomly with replacement from the short trajectories (i.e., from all possible such pairs in the data set).
{
From each frame, we use only the atom positions because the momenta decorrelate within a few picoseconds, which is much shorter than the lag times that we consider.  However, in principle, the momenta could impact the prediction functions \cite{metzner2006illustration} and be used as neural-network inputs as well.
}



\section{Spectral estimation}\label{sec:eig}

In this section, we provide some further numerical details for the application of our method to spectral estimation and demonstrate the method on the test problems.  For our subspace iteration with ${\cal A}={\cal T}^\tau$, we require estimators for inner products of the form $\langle f, \T{\tau} g\rangle_\mu$.
For example, the gradient of the loss function \eqref{eq:general_Loss} involves inner products of the form
\begin{equation}
\left\langle\nabla_\theta \varphi_\theta^a,  \mathcal{T}^\tau \varphi_{\theta}^b\right\rangle_\mu.
\end{equation}
For these, we use the unbiased data-driven estimator
\begin{equation}
\langle f, \T{\tau} g\rangle_\mu \approx \frac{1}{n} \sum_{j=1}^n  f(X_j^0) g(X_j^\tau).
\end{equation}

As discussed in Section~\ref{sec:ieSI}, applying the operator $\mathcal{T}^\tau$ repeatedly causes each basis function to converge to the dominant eigenfunction and leads to numerical instabilities.  To avoid this, we orthogonalize the outputs of the networks with a QR decomposition at the end of each subspace iteration by constructing the matrix 
$\Phi_{ia} = \varphi^a_{\theta}(X_i^0)$
and then computing the factorization $\Phi =QR$ where $Q$ is an $n\times k$ matrix with orthogonal columns and $R$ is an upper triangular $k\times k$ matrix.  Finally, we set $\tilde{\varphi}_s^a= \sum_{b=1}^k\varphi_\theta^b\,(R^{-1} N)_{ba}$, where $N$ is a diagonal matrix with entries equal to the norms of the columns of $\Phi$ (before orthogonalization).
To ensure that the networks remain well-separated (i.e., the eigenvalues of $C^0$ remain away from zero), we penalize large off-diagonal entries of $K$ by adding to the loss
\begin{equation}
\label{eq:K_reg}
\gamma_1{\norm{K-\mathrm{diag}(K)}}^2_\textrm{F},
\end{equation}
where $\gamma_1$ allows us to tune the strength of this term relative to others, and $\norm{\cdot}_\textrm{F}$ is the Frobenius norm. 
We control the scale of each network output using the strategy from Ref.~\onlinecite{wen_batch_2020};  that is, we add to the loss a term of the form
\begin{equation}
\gamma_2\sum_{a=1}^k\left[2\nu_a\left(\frac{1}{n}\sum_{j=1}^n\varphi_\theta^a(X_j^0)^2-1\right)-\nu_a^2\right],
\end{equation}
where we have introduced the conjugate variables $\nu_a$ which we maximize with gradient ascent (or similar optimization).
In general, our numerical experiments suggest that it is best to keep $\gamma_1$ and $\gamma_2$ relatively small.  We find that stability of the algorithm over many subspace iterations is improved if the matrix $K$ is set at its optimal value before each inner loop.  To do this, we set
\begin{equation}
\begin{aligned}
K_{1:i,i}=\arg \min_{c}\frac{1}{n}&\sum_{j=1}^n\left(\sum_{a=1}^i\varphi_{\theta}^{a}(X_j^0)c_a-\tilde{\varphi}_s^a(X_j^{\tau})\right)^2 \\
&+\gamma_2\sum_{a=1}^{i-1}c_a^2.
\end{aligned}
\end{equation}
The above minimization can be solved with linear least squares.
Finally, we note that in practice any optimizer can be used for the inner iteration steps, though the algorithm below implements stochastic gradient descent.  In this work, we use Adam \cite{kingma2017adam} for all numerical tests.
We summarize our procedure for spectral estimation in Algorithm~\ref{alg:eig}.

\begin{figure*}
\begin{minipage}{\linewidth}
\begin{algorithm}[H]
\caption{Inexact subspace iteration (with $L^2_\mu$ loss) for spectral estimation}
\label{alg:eig}
\begin{algorithmic}[1]
\Require{
Subspace dimension $k$,
transition data $\{X^0_j,X^{\tau}_j\}_{j=1}^n$,
batch size $B$, learning rate $\eta$, number of subspace iterations $S$, number of inner iterations $M,$ regularization parameters $\gamma_1$ and $\gamma_2$}
\State Initialize $\{\varphi_\theta^a\}_{a=1}^k$ and $\set{\tilde \varphi_1^a}_{a=1}^k$
\For{$s = 1\dots S$}
    \For{$m = 1\dots M$}
        \State Sample a batch of data $\{X^0_j,X^{\tau}_j\}_{j=1}^B$
        \State $\hat{\mathcal{L}}_1 \gets \frac{1}{B} \sum_{j=1}^B \sum_{a=1}^k\left[\frac{1}{2} (\sum_{b=1}^k\varphi_\theta^{b}(X_j^0)K_{b a})^2
        -\sum_{b=1}^k\varphi_\theta^b(X_j^0)K_{ba}\left\{ \alpha_s \tilde{\varphi}_s^a(X_j^{\tau}) + (1 - \alpha_s) \tilde{\varphi}_s^a(X_j^0)\right\}\right]$ 
        \State $\hat{\mathcal{L}}_{K} \gets \gamma_1{\norm{K-\mathrm{diag}(K)}}^2_\textrm{F}$
        \State $\hat{\mathcal{L}}_{\rm{norm}} \gets \gamma_2\sum_{a=1}^k(2\nu_a( \frac{1}{B}\sum_{j=1}^B(\varphi_\theta^a(X_j^0)^2)-1)-\nu_a^2)$
        \State $\hat{\mathcal{L}} \gets \hat{\mathcal{L}}_1 + \hat{\mathcal{L}}_K + \hat{\mathcal{L}}_{\rm{norm}}$
        \State $\theta \gets \theta-\eta \nabla_{\theta}\hat{\mathcal{L}}$
        \State $K\gets K-\eta\, \mathrm{triu}(\nabla_K\hat{\mathcal{L}})$
        \State $\nu_a \gets \nu_a+\eta\nabla_{\nu_a}\hat{\mathcal{L}}$
    \EndFor
    \State Compute the matrix $\Phi_{ia} = \varphi^a_{\theta}(X_i^0)$ 
    \Comment{$\Phi \in \mathbb{R}^{n\times k}$}
    \State Compute diagonal matrix $N^2_{aa}=\sum_i\varphi^a_{\theta}(X_i^0)^2$
    \State Compute QR-decomposition $\Phi = QR$ \Comment{$Q\in \mathbb{R}^{n\times 
    k}$ and $R\in \mathbb{R}^{k\times k}$}
    \State $\tilde{\varphi}^a_{s} \gets \sum_{b=1}^k \varphi^b_{\theta}\,(R^{-1}N)_{ba}$
    \State $K_{1:i,i}\gets\arg \min_{c}\frac{1}{n}\sum_{j=1}^n\left(\sum_{a=1}^i\varphi_{\theta}^{a}(X_j^0)c_a-\tilde{\varphi}_s^a(X_j^{\tau})\right)^2+\gamma_2\sum_{a=1}^{i-1}c_a^2 $
\EndFor
\State Compute the matrices $C^t_{a b}=\frac{1}{n}\sum_{j=1}^n\tilde \varphi_s^a(X_j^0)\tilde \varphi_s^b(X_j^t)$ for $t=0,\tau$ \Comment{$C^t \in \mathbb{R}^{k\times k}$}
\State Solve the generalized eigenproblem $C^\tau W=C^0 W\Lambda$ for $W$ and $\Lambda$ 
\State \Return eigenvalues $\Lambda$, eigenfunctions $v_a=\sum_{b=1}^kW_{ab}\tilde \varphi_s^b$ 
\end{algorithmic}
\end{algorithm}
\end{minipage}
\end{figure*}

\subsection{M\"uller-Brown model}\label{sec:mbeigenfunctions}

\begin{table*}[htb]
    \centering
    \caption{Parameter choices used in this work}
    \begin{tabular}{c||cc|ccc|c}
     \hline
        & \multicolumn{2}{c|}{\textbf{Spectral Estimation}} & \multicolumn{3}{c|}{\textbf{Committor}} & \textbf{MFPT} \\ \hline
         \textbf{Hyperparameter} & M\"uller-Brown & AIB\textsubscript{9} & M\"uller-Brown & Modified M\"uller-Brown & AIB\textsubscript{9} & AIB\textsubscript{9} \\ \hline
         subspace dimension $k$ & 3 & 5 & 1 & 2, 1\footnote{Four subspace iterations with $k=2$ followed by ten iterations with $k=1$} & 1 & 5 \\
         input dimensionality & 2 & 174 & 2 & 2 & 174 & 174\\
         hidden layers & 6 & 6 & 6 & 6 & 6 & 6\\
         hidden layer width & 64 & 128 & 64 & 64 & 150 & 150\\
         hidden layer nonlinearity & CeLU & CeLU & ReLU & ReLU & ReLU & ReLU \\
         output layer nonlinearity & none & tanh & sigmoid/none & none & none & none \\
         outer iterations $S$ & 10 & 100 & 100 & 4 + 10\footnotemark[1] & 100 & 300 \\ 
         inner iterations $M$ & 5000 & 2000 & 200 & 5000 & 2000 & 1000\\ 
         $\sigma$ & 2 & 50 & 50 & 0 & 50 & 0\\ 
         batch size $B$ & 2000 & 1024 & 5000 & 2000 & 1024 & 2000 \\ 
         learning rate $\eta$ & 0.001 & 0.0001 & 0.001 & 0.001 & 0.001 & 0.001\\ 
         $\gamma_1$ & 0.15 & 0.001 & - & - & - & 0.1\\ 
         $\gamma_2$ & 0.01 & 0.01 & - & - & - & 10\\ 
        loss for  $\varphi_\theta^1$ & $L^2_\mu$ & $L^2_\mu$ & $L^2_\mu$/softplus & softplus & softplus & $L^2_\mu$\\ 
        loss for $\varphi_\theta^a$ for $a > 1$ & $L^2_\mu$ & $L^2_\mu$ & - & $L^2_\mu$ & - & $L^2_\mu$\\
        \hline
    \end{tabular}
    \label{tab:params}
\end{table*}

As a first test of our method, we compute the $k=3$ dominant eigenpairs for the M\"uller-Brown model.  
Since we know that the dominant eigenfunction of the transition operator is the constant function $v_1(y,z) = 1$ with eigenvalue $\lambda_1 = 1$, we directly include this function in the basis as a non-trainable function, i.e. $\varphi_\theta^1(y,z)=1.$
To initialize $\tilde{\varphi}_1^a$ for each $a>1$, we choose a standard Gaussian vector $(Y^a,Z^a)\in \mathbb{R}^2$, and set $\tilde{\varphi}_1^a(y,z)=y\, Y^a + z\,Z^a.$  This ensures that the initial basis vectors are well-separated and the first QR step is numerically stable. 
 Here and in all subsequent M\"uller-Brown tests, batches of trajectories are drawn from the entire data set with replacement. Other hyperparameters are listed in Table~\ref{tab:params}.

\begin{figure}[hbt]
\begin{center}
    \includegraphics[width=\linewidth]{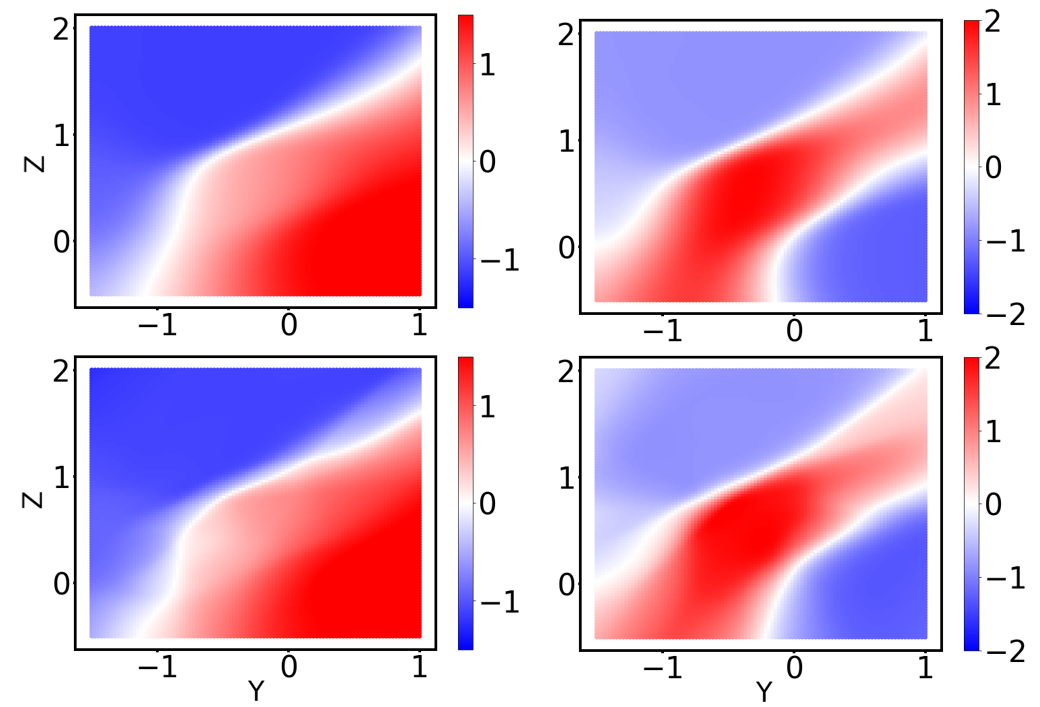}
\end{center}
    \caption{First two non-trivial eigenfunctions of the M\"uller-Brown model.  (top) Grid-based reference. (bottom) Neural network subspace after ten subspace iteration steps, computed with $\tau=300$ steps (i.e., $0.3\,\delta_t^{-1}$). 
    }
    \label{fig:mb_eigfn}
\end{figure}

\begin{figure*}[bt]
\centering
    \includegraphics[width=0.8\textwidth]
    {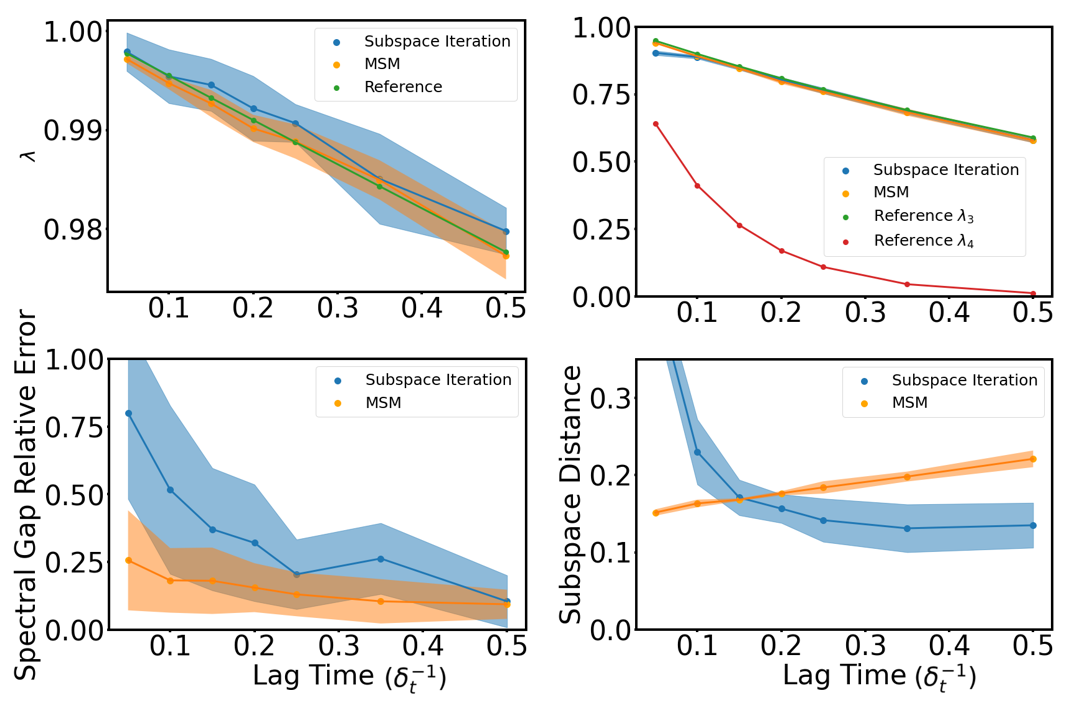}
    \caption{
    Spectral estimation as a function of lag time (in units of $\delta_t^{-1}$) for the M\"uller-Brown model.
    (top left) Second eigenvalue.  (top right) Third and fourth eigenvalues; only the reference fourth eigenvalue is shown to illustrate the spectral gap.  (bottom left) Relative error in the first spectral gap (i.e., $1-\lambda_2$).  (bottom right) Subspace distance between estimated and reference three-dimensional invariant subspaces.
    }
    \label{fig:mb_subdist}
\end{figure*}

Figure~\ref{fig:mb_eigfn} shows that we obtain good agreement between the estimate produced by the inexact subspace iteration in Algorithm \ref{alg:eig} and reference eigenfunctions.   Figure~\ref{fig:mb_subdist} (upper panels) shows how the corresponding eigenvalues vary with lag time; again there is good agreement with the reference. Furthermore, there is a significant gap between $\lambda_3$ and $\lambda_4$, indicating that a three-dimensional subspace captures the dynamics of interest for this system.

We compare the subspace that we obtain from our method with that from an MSM constructed 
{
from the same amount of data by using $k$-means to cluster the configurations into 400 states and counting the transitions between clusters.} This is a very fine discretization for this system, and the MSM is sufficiently expressive to yield eigenfunctions in good agreement with the reference. The relative error of $1 -\lambda_2$ is comparable for the two methods (Figure \ref{fig:mb_subdist}, lower left).
To compare two finite dimensional subspaces,  $\mathcal{U}$ and $\mathcal{V}$, 
we define the subspace distance as \cite{webber2021error}
\begin{equation}
    d(\mathcal{U}, \mathcal{V}) = \norm[\big]{(I-P_{\mathcal{U}})P_{\mathcal{V}}}_\textrm{F},
\end{equation}
where $P_{\mathcal{U}}$ and ${P}_{\mathcal{V}}$ denote the orthogonal projections  onto $\mathcal{U}$ and $\mathcal{V}$, respectively, and $\norm{\cdot}_\textrm{F}$ is the Frobenius norm.
Figure \ref{fig:mb_subdist} (lower right) shows the subspace distances from the reference as functions of lag time. We see that the inexact subspace iteration better approximates the three-dimensional dominant eigenspace for moderate to long lag times, even though the eigenvalues are comparable.

\subsection{\texorpdfstring{AIB\textsubscript{9}}{AIB9}}\label{sec:eig_aib9}
For the molecular test system, we compute the dominant five-dimensional subspace.
We compare the inexact subspace iteration in Algorithm \ref{alg:eig} with MSMs constructed on dihedral angles (``dihedral MSM'') and on Cartesian coordinates (``Cartesian MSM'').
We expect the dihedral MSM to be accurate given that the dynamics of AIB\textsubscript{9} are well-described by the backbone dihedral angles \cite{buchenberg_hierarchical_2015, perez_meld-path_2018}, and we thus use it as a reference.
It is constructed by clustering the sine and cosine of each of the backbone dihedral angles ($\phi$ and $\psi$) for the nine residues (for a total of $2 \times2\times 9 = 36$ input features) into 1000 clusters using $k$-means { and counting transitions between clusters.
The Cartesian MSM is constructed by similarly counting transitions between 1000 clusters from the $k$-means algorithm, but the clusters are }based on the Cartesian coordinates of all non-hydrogen atoms after aligning the backbone atoms of the trajectories, for a total of 174 input features.
Because of the difficulty of clustering high-dimensional data, we expect the Cartesian MSM basis to give poor results.
The neural network for the inexact subspace iteration receives the same 174 Cartesian coordinates as input features.  We choose to use Cartesian coordinates rather than dihedral angles as inputs because it requires the network to identify nontrivial representations for describing the dynamics.

As in the M\"uller-Brown example, we use ${\varphi}_\theta^1=1$ and  a random linear combination of coordinate functions to initialize $\tilde{\varphi}_1^a$ for $a>1$. 
Other hyperparameters are listed in Table~\ref{tab:params}.  
{
With these choices, the neural-network training typically requires about 20 minutes on a single NVIDIA A40 GPU; this is much longer than the time required for diagonalization of the $1000\times 1000$ MSM transition matrix, which is nearly instantaneous.  However, the time for neural-network training is negligible compared with the time to generate the data set, which is the same for both approaches.
}

\begin{figure*}[hbt]
    \centering
    \includegraphics[width=0.8\textwidth]{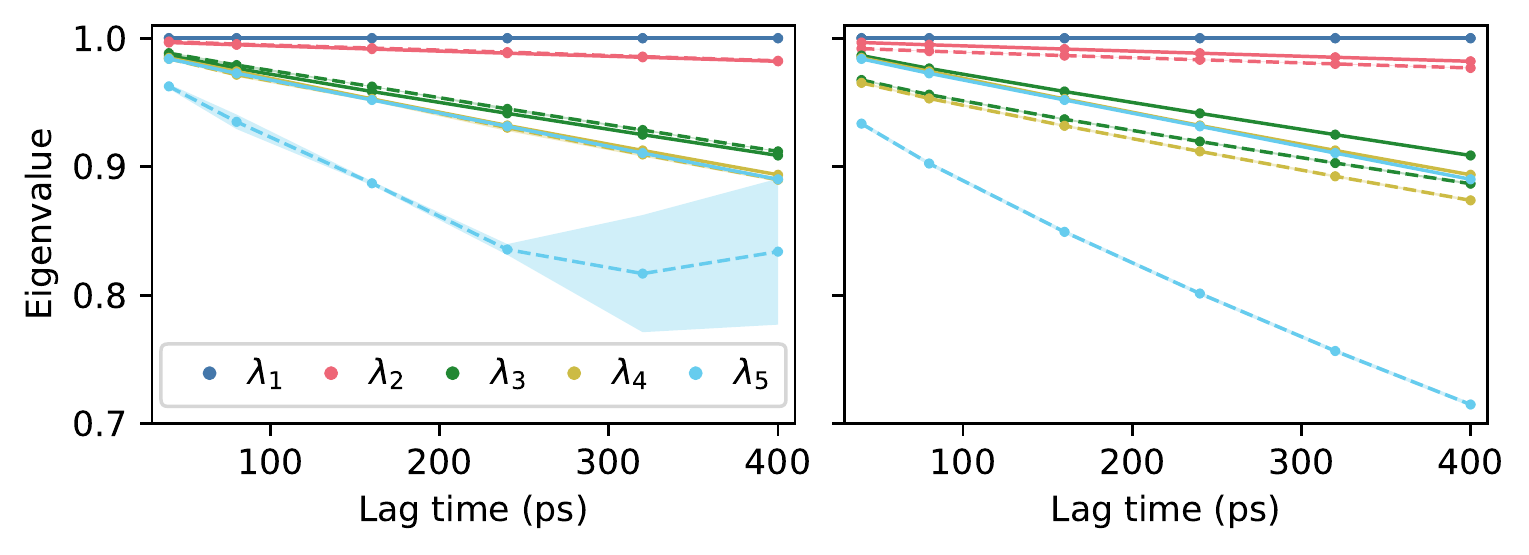}
    \caption{First five eigenvalues of the transition operator for AIB\textsubscript{9} as a function of lag time.
    (left) Comparison between eigenvalues computed using the dihedral MSM with 1000 clusters (solid lines) and the inexact subspace iteration (dashed lines). The shading indicates standard deviations over five trained networks for the subspace iteration. 
    (right) Comparison between a dihedral MSM (solid lines) and Cartesian MSMs with 1000 clusters (dashed lines). The standard deviations for the Cartesian MSMs over five random seeds for $k$-means clustering are too narrow to be seen.
    }
    \label{fig:aib9_evals}
\end{figure*}

\begin{figure}[hbt]
    \centering
    \includegraphics[width=\linewidth]{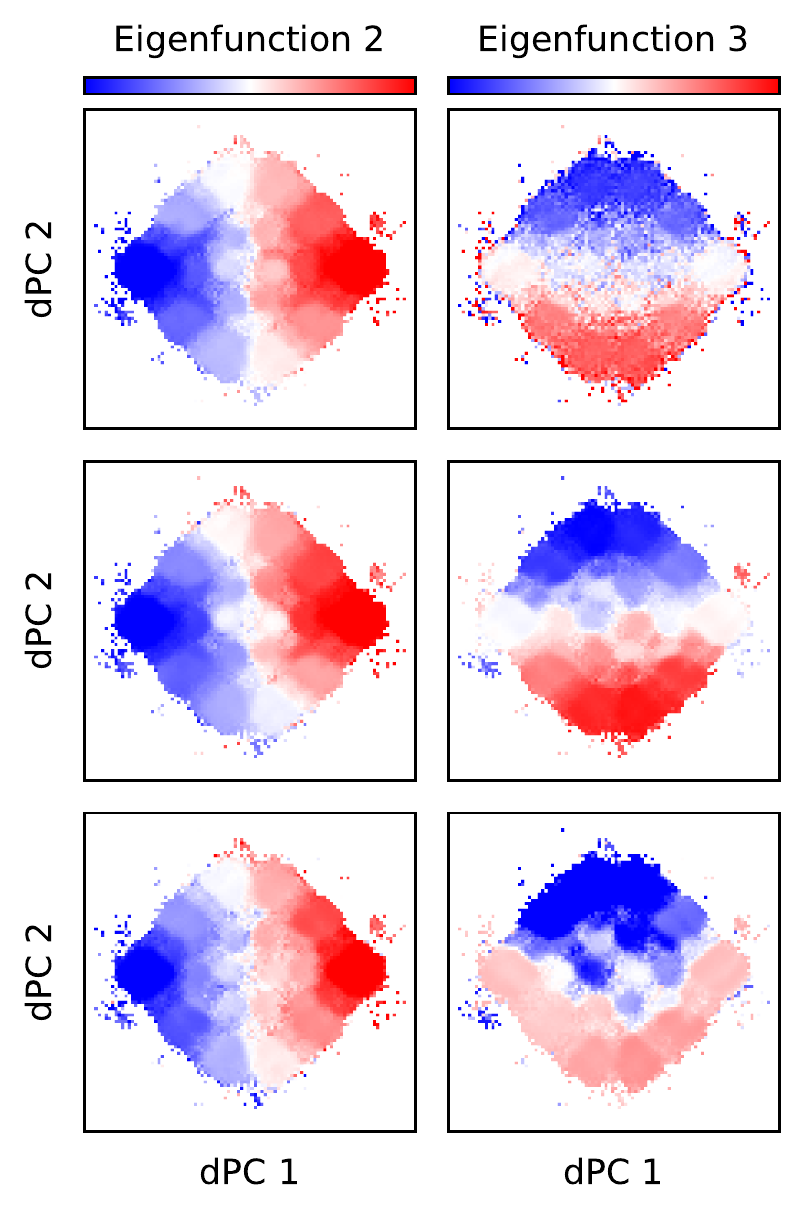}
    \caption{First two non-trivial eigenfunctions of AIB\textsubscript{9} projected onto the first two dPCs (i.e., averaged for bins in the two-dimensional space shown). (top) MSM constructed on sine and cosine of dihedral angles with 1000 clusters and  lag time corresponding to 40 ps. (middle) Inexact subspace iteration using Cartesian coordinates and the same lag time.
    (bottom) MSM constructed on Cartesian coordinates with 1000 clusters and the same lag time.}
    \label{fig:aib9_eigfn}
\end{figure}

Taking the dihedral MSM as a reference, the Cartesian MSM systematically underestimates the eigenvalues (Figure~\ref{fig:aib9_evals}).
The subspace iteration is very accurate for the first four eigenvalues but the estimates for the fifth are low and vary considerably from run to run.
A very small gap between $\lambda_4$ and $\lambda_5$ may contribute to the difficulty in estimating $\lambda_5$.
In Figure~\ref{fig:aib9_eigfn}, we plot the first two non-trivial eigenfunctions ($v_2$ and $v_3$), which align with the axes of the dPC projection.
The eigenfunction $v_2$ corresponds to the transition between the left- and right-handed helices; the eigenfunction $v_3$ is nearly orthogonal to $v_2$ and corresponds to transitions between intermediate states.
It is challenging to visualize the remaining two eigenfunctions by projecting onto the first two dPCs because $v_4$ and $v_5$ are orthogonal to $v_2$ and $v_3$.
The estimates for $v_2$ are in qualitative agreement for all lag times tested (Figure \ref{fig:aib9_eigfn} shows results for $\tau$ corresponding to 40 ps), but the subspace iteration results are less noisy for the shortest lag times.
Moreover, the estimate for $v_3$ from subspace iteration agrees more closely with that from the dihedral MSM than does the estimate for $v_3$ from the Cartesian MSM.
The subspace distance for $v_2$ and $v_3$ between the subspace iteration and the dihedral MSM is 0.947, compared with a value of 0.969 for the subspace distance between the two MSMs.
Together, our results indicate that the neural networks are able to learn the leading eigenfunctions and eigenvalues of the transition operator (dynamical modes) of this system despite being presented with coordinates that are not the natural ones for describing the dynamics.

\section{Prediction}\label{sec:forecast}

Inexact subspace iteration for $\mathcal{A}$ in \eqref{eq:A1} is equivalent to performing the inexact Richardson iteration in \eqref{eq:ierich} on the first basis function $\varphi_\theta^1$ and then performing an inexact subspace iteration for the operator $\mathcal{S}^\tau$ on the rest of the basis functions. The iteration requires unbiased estimators of the forms
\begin{equation}
\inner{ f }{ \sT{\tau} g}_\mu \approx \frac{1}{n} \sum_{j=1}^n  f(X_j^0) g(X_j^{\tau\bmin T_j})
\end{equation}
and
\begin{equation}
\inner{ f }{ r }_\mu \approx \frac{1}{n} \sum_{j=1}^n  f(X_j^0) \sum_{t=0}^{(\tau\bmin T_j)-1} \Gamma(X_j^t),
\end{equation}
where $T_j$ is the first time $X_j^t$ enters $D^\comp$ and $r$ is the right-hand side of \eqref{eq:FK}, as previously.

The Richardson iterate, $\varphi_\theta^1$, must satisfy the boundary condition $\varphi_\theta^1(x) = \Psi(x)$ for $x\notin D$. The other basis functions should satisfy $\varphi^a_\theta(x) = 0$ for $x\notin D$.  
In practice,  we enforce the boundary conditions by explicitly setting  $\varphi^1_\theta(x)=\Psi(x)$  and $\varphi^a_\theta(x)=0$ for $a>1$ when  $x\notin D$.

When the boundary condition is zero, as for the MFPT, we find an approximate solution of the form 
\begin{equation}
 u_{\theta} =  \sum_{a=1}^k w_a \varphi^a_{\theta} 
 \end{equation}
 by 
solving the $k$-dimensional linear system
\begin{equation}\label{eq:linsys}
\left(C^0 - C^\tau\right) w = p
\end{equation}
where, for $a, b \geq 1$,
\begin{equation}\label{eq:C}
C^t_{ab} = \inner{\varphi^a_{\theta}}{ \sT{t} \varphi^b_{\theta}}_\mu 
\end{equation}
for $t = 0, \tau$, and
\begin{equation}\label{eq:v}
    p_a = \inner[\big]{ \varphi^a_{\theta}}{\E_x\left[ \rho(X) \right]  }_\mu.
\end{equation}
In \eqref{eq:v}, we introduce the notation
 \begin{equation}
\rho(X) =\sum_{t=0}^{(\tau\bmin T)-1} \Gamma(X^t) 
\end{equation}
for use in Algorithm \ref{alg:forecast}.

When the boundary condition is non-zero, as for the committor, 
we restrict \eqref{eq:linsys} to a  $(k-1)$-dimensional linear system by excluding the indices $a=1$ and $b=1$ in \eqref{eq:C} and \eqref{eq:v} 
and setting
 \begin{equation}
\rho(X) = \varphi^1_{\theta}(X^{\tau\wedge T}) 
 - \varphi^1_{\theta}(X^0)
 +\sum_{t=0}^{(\tau\bmin T)-1} \Gamma(X^t).
\end{equation}
In this case the corresponding approximate solution is 
 \begin{equation}
u_{\theta} = \varphi^1_{\theta} + \sum_{a=2}^k w_a \varphi^a_{\theta}.
\end{equation}
This approximate solution corresponds to the one given by dynamical Galerkin approximation \cite{thiede_galerkin_2019, strahan_long-time-scale_2021} with the basis ${\{\varphi^a_{\theta}\}}_{a=2}^k$ and a ``guess'' function of $\varphi^1_{\theta}$.

When the boundary conditions are zero, the orthogonalization procedure and the matrix $K$ are applied to all basis functions as in Section \ref{sec:eig}.
When the boundary conditions are non-zero, the orthogonalization procedure is only applied to the basis functions ${\{\varphi^a_{\theta}\}}_{a=2}^k$, and $K_{a 1} = I_{a 1}$,  the $a1$ element of the identity matrix. 
We summarize our procedure for prediction in Algorithm \ref{alg:forecast}.

\begin{figure}[bt]
\begin{center}
    \includegraphics[width=0.9\linewidth]{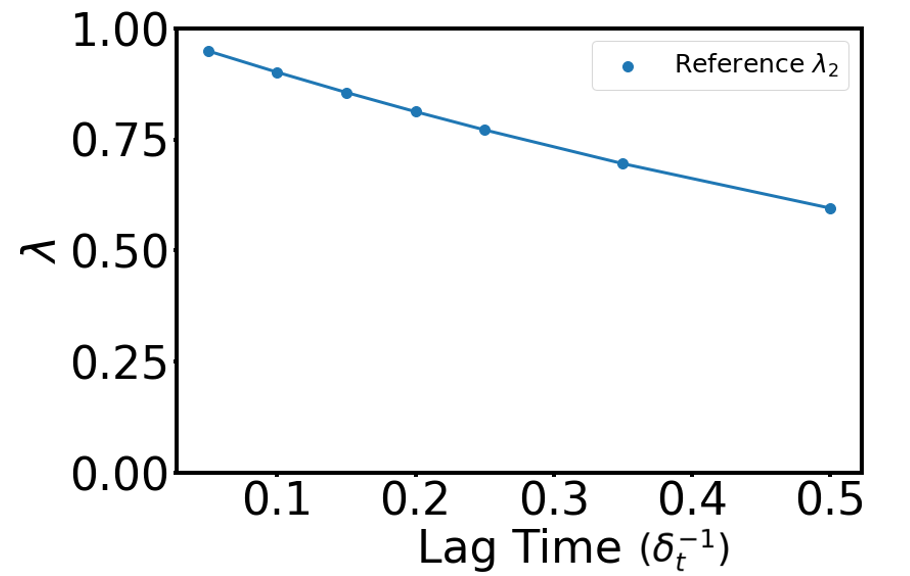}
\end{center}
    \caption{
    First eigenvalue of $\mathcal{S}^\tau$ (second of  $\mathcal{A}$ in \eqref{eq:A1}) for the M\"uller-Brown model as a function of lag time (in units of $\delta_t^{-1}$). The gap between this eigenvalue and the dominant eigenvalue, which is one, determines the rate of convergence of the Richardson iteration. 
    }
    \label{fig:MB_q_secondeigenvalue}
\end{figure}

\begin{figure*}
\begin{minipage}{\linewidth}
\begin{algorithm}[H]
\caption{Inexact subspace iteration (with $L^2_\mu$ loss) for prediction functions}
\label{alg:forecast}
\begin{algorithmic}[1]
\Require{Subspace dimension $k$, stopped transition data ${\set{X^0_j,X^{\tau \bmin T_j}_j }}_{j=1}^n$, reward data ${\set{ \rho(X_j)}}_{j=1}^n$, batch size $B$, learning rate $\eta$, number of subspace iterations $S$, number of inner iterations $M,$ regularization parameters $\gamma_1$ and $\gamma_2$}
\State Initialize $\{\varphi_\theta^a\}_{a=1}^k$ and $\set{\tilde \varphi_1^a}_{a=1}^k$
\For{$s = 1\dots S$}
    \For{$m = 1\dots M$}
        \State Sample a batch of data $\{X^0_j,X^{\tau\bmin T_j}_j\}_{j=1}^B$, $\set{\rho(X_j)}_{j=1}^B$
        \State $\hat{\mathcal{L}}_1 \gets \frac{1}{B} \sum_{j=1}^B \sum_{a=1}^k \left[ \frac{1}{2} (\sum_{b=1}^k\varphi_\theta^{b}(X_j^0)K_{b a} )^2 
        - \alpha_s\left(\sum_{b=1}^k \varphi_\theta^b K_{ba} (\tilde{\varphi}_s^a(X_j^{\tau \bmin T_j})+\rho(X_j)I_{a 1} ) \right) \right]$
        \State $\hat{\mathcal{L}}_2 \gets - \frac{1}{B} \sum_{j=1}^B \sum_{a=1}^k (1 - \alpha_s) \sum_{b=1}^k \varphi_\theta^b K_{ba} \tilde{\varphi}_s^a(X_j^0)$
        \State $\hat{\mathcal{L}}_K \gets \gamma_1{\norm{K-\mathrm{diag}(K)}}^2_F$
        \State $\hat{\mathcal{L}}_{\rm{norm}} \gets \gamma_2\sum_{a=2}^k(2\nu_a( \frac{1}{B}\sum_{j=1}^B(\varphi_\theta^a(X_j^0)^2)-1)-\nu_a^2)$
        \State $\hat{\mathcal{L}} \gets \hat{\mathcal{L}}_1 +\hat{\mathcal{L}}_2 +  \hat{\mathcal{L}}_K + \hat{\mathcal{L}}_{\rm{norm}}$
        \State $\theta \gets \theta-\eta \nabla_\theta \hat {\mathcal{L}}$
        \State $K \gets K-\eta \,(\mathrm{triu}(\nabla_K \hat {\mathcal{L}}))$
        \State $\nu_a \gets \nu_a+\eta \nabla_{\nu_a} \hat {\mathcal{L}}$
    \EndFor
    \If{$\Psi(x)=0$}
        \State Compute the matrix $\Phi_{ia} = \varphi^a_\theta (X_i^0)$ \Comment{$\Phi \in \mathbb{R}^{n\times k}$}
    \Else
        \State Compute the matrix $\Phi_{ia} = \varphi^a_\theta(X_i^0)$ for $a > 1$ \Comment{$\Phi \in \mathbb{R}^{n\times (k-1)}$}
        
    \EndIf
    \State Compute QR-decomposition $\Phi = QR$ 
    \State Compute diagonal matrix $N^2_{aa}=\sum_i\varphi^a_{\theta}(X_i^0)^2$
    \State $\tilde{\varphi}^a_{s} \gets \sum_{b=1}^k \varphi^b_{\theta}\, (R^{-1}N)_{ba}$ 
    \Comment{if $\Psi(x) = 0$ exclude $a = 1$}
    \State $K_{1:i,i}\gets\arg \min_{c}\frac{1}{n}\sum_{j=1}^n\left(\sum_{a=1}^i\varphi_{\theta}^{a}(X_j^0)c_a-\tilde{\varphi}_s^a(X_j^{\tau \bmin T_j})\right)^2+\gamma_2\sum_{a=1}^{i-1}c_a^2 $
\EndFor
\State Compute the matrix $C^t_{a b}=\frac{1}{n}\sum_{j=1}^n \varphi^a_\theta (X_j^0) \varphi^b_\theta (X_j^{t})$ for $t=0,\tau \bmin T_j$ \Comment{$C^t\in \mathbb{R}^{k\times k}$}
\State Compute the vector $p_a= \frac{1}{n}\sum_{j=1}^n \varphi^a_\theta (X_j^0)\rho(X_j)$
\State Solve linear system $(C^0-C^\tau)w=p$ \Comment{if $\Psi(x) = 0$ enforce $w_1 = 1$}
\State \Return $u=\sum_{a=1}^k w_a \varphi_\theta^a$ 
\end{algorithmic}
\end{algorithm}
\end{minipage}
\end{figure*}

\subsection{M\"uller-Brown committor}\label{sec:mbcommittor}

In this section, we demonstrate the use of our method for prediction by estimating the committor for the M\"uller-Brown model with a shallow intermediate basin at $(-0.25,0.5)$ (Figure~\ref{fig:MB}).   Here the sets $A$ and $B$ are defined as in Eq.~\eqref{eqn:MBstates} and $T$ is the time of first entrance to $D^\comp = A\cup B$. In this case, a one-dimensional subspace iteration (i.e., $k=1$ in Algorithm \ref{alg:forecast}) appears sufficient to accurately solve the prediction problem.  Figure \ref{fig:MB_q_secondeigenvalue} shows the largest eigenvalue of the { stopped transition operator $\mathcal{S}^\tau$ (the second largest of $\mathcal{A}$ in \eqref{eq:A1}) computed} from our grid-based reference scheme (Section \ref{sec:mb_numerics}).
Richardson iteration should converge geometrically in this eigenvalue \cite{GoluVanl96}, and so, for moderate lag times, we can expect our method to converge in a few dozen iterations. To initialize the algorithm we choose $\tilde{\varphi}_1^1=\ind{B}$.  
All other hyperparameters are listed in Table~\ref{tab:params}.

\begin{figure*}[bt]
\begin{center}
    \includegraphics[width=0.8\textwidth]{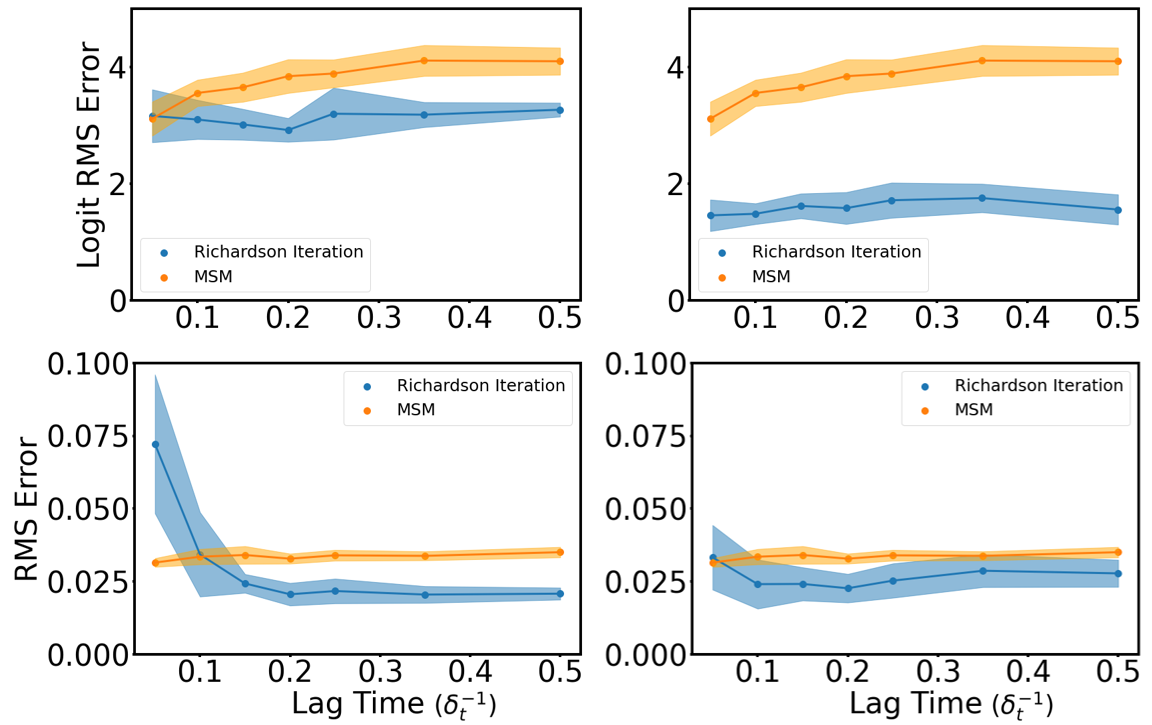}
\end{center}
    \caption{
    Committor prediction for the M\"uller-Brown system as a function of lag time (in units of $\delta_t^{-1}$).
    (left) Comparison of the inexact Richardson iteration using the $L^2_\mu$ loss  and an MSM with 400 states. (right) Same comparison using the softplus loss in place of the $L^2_\mu$ loss.
    }
    \label{fig:MB_q}
\end{figure*}

\begin{figure*}[bt]
\begin{center}
    \includegraphics[width=0.8\textwidth]{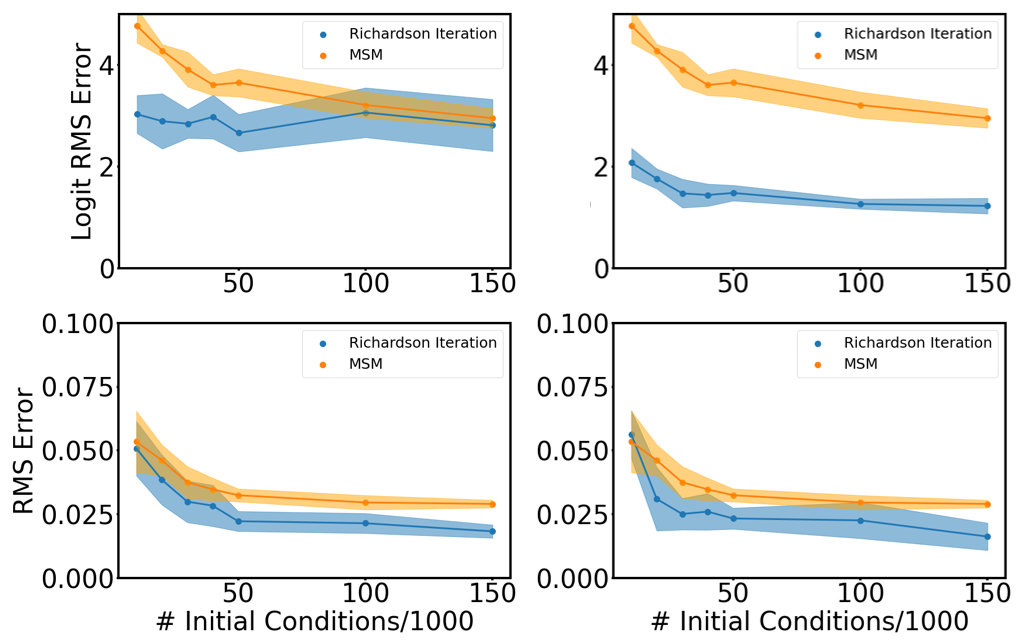}
\end{center}
    \caption{
    Committor prediction for the M\"uller-Brown as a function of number of initial conditions for a fixed lag time of $\tau = 0.1 \delta_t^{-1}$. (left) Comparison of inexact Richardson iteration using the $L^2_\mu$ loss  and an MSM with 400 states. (right) Same comparison using the softplus loss in place of the $L^2_\mu$ loss.
    }
    \label{fig:MB_q_Data}
\end{figure*}


We compare the estimate of the committor from our approach with that from an MSM { constructed from the same amount of data by using $k$-means to cluster the configurations outside $A$ and $B$ into 400 states and counting the transitions between clusters}.
In addition to the root mean square error (RMSE) for the committor itself, we show the RMSE of  
\begin{equation}\label{eq:logit}
\mathrm{logit}_{\varepsilon}(q)=\log\left(\frac{\varepsilon+q}{1+\varepsilon-q}\right)
\end{equation}
for points outside $A$ and $B$.  This function amplifies the importance of values close to zero and one.
We include $\varepsilon$ because we want to assign only a finite penalty if the procedure estimates $q$ to be exactly zero or one; we use $\varepsilon=e^{-20}$.

Results as a function of lag time are shown in Figure \ref{fig:MB_q}.  We see that the Richardson iterate is more accurate than the MSM for all but the shortest lag times.  When using the $L^2_\mu$ loss, the results are comparable, whereas the softplus loss allows the Richardson iterate to improve the RMSE of the logit function in \eqref{eq:logit} with no decrease in performance with respect to the RMSE of the committor.  Results as a function of the size of the data set are shown in Figure \ref{fig:MB_q_Data} for a fixed lag time of $\tau = 0.1 \delta_t^{-1}$.  The Richardson iterate generally does as well or better than the MSM. Again, the differences are more apparent in the RMSE of the logit function in \eqref{eq:logit}.  By that measure, the Richardson iterate obtained with both loss functions is significantly more accurate than the MSM for small numbers of trajectories. The softplus loss maintains an advantage even for large numbers of trajectories.

\subsection{Accelerating convergence by incorporating eigenfunctions}\label{sec:Mod_MB}

\begin{figure*}[bt]
    \centering
    \includegraphics[width=0.8\textwidth]{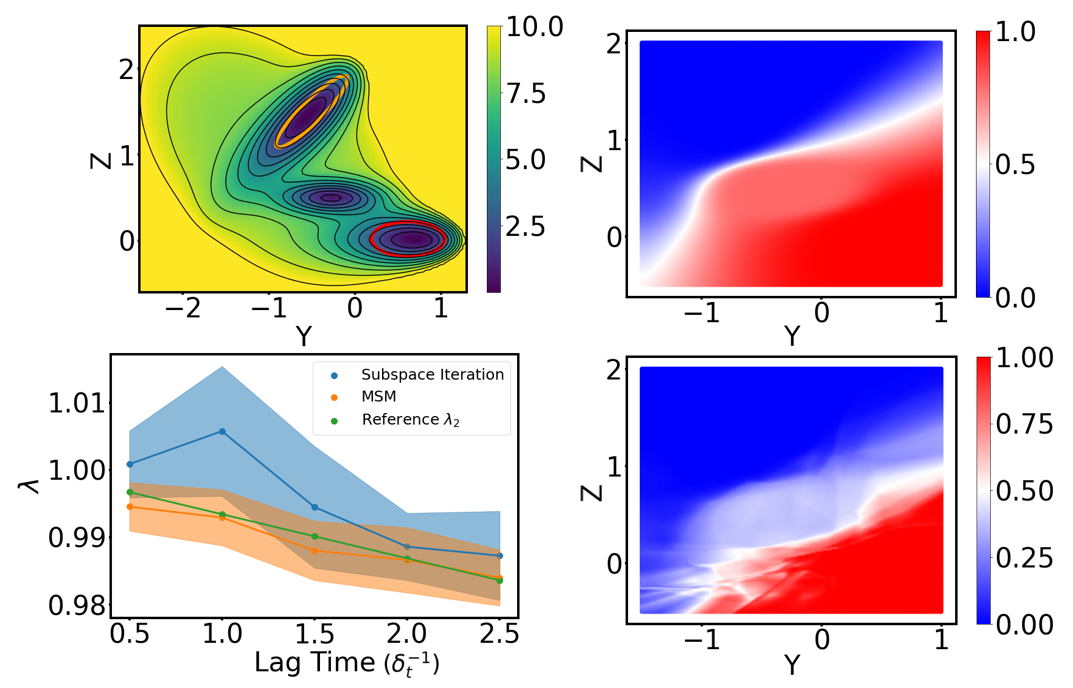}
    \caption{Richardson iteration for the committor converges slowly for a M\"uller-Brown potential with a deepened intermediate.  (top left) Potential energy surface, with states $A$ and $B$ indicated. Contour lines are drawn every 1 $\beta^{-1}$.  (top right) Reference committor. (bottom left) Dominant eigenvalue as a function of lag time (in units of $\delta_t^{-1}$) from an MSM with 400 states, subspace iteration, and the grid-based reference. 
    (bottom right) Example of the Richardson iteration after 400 iterations.  Note the overfitting artifacts and lack of convergence near the intermediate state.}
    \label{fig:SI_Intro}
\end{figure*}

As discussed in Section~\ref{sec:inexactPower},  we expect  Richardson iteration to converge slowly when the largest eigenvalue of $\mathcal{S}^\tau$, $\lambda_1$, is close to 1.  More precisely, the number of iterations required to reach convergence should scale with $-1/\log \lambda_1=\mathbb{E}\left[T\right]/\tau$,  the mean escape time from the quasi-stationary distribution to the boundary of $D$ divided by the lag time.  With this in mind, we can expect inexact Richardson iteration for the M\"uller-Brown to perform poorly if we deepen the intermediate basin at $(-0.25,0.5)$ as in Figure~\ref{fig:SI_Intro} (top left). 
Again, the sets $A$ and $B$ are defined as in \eqref{eqn:MBstates}, and $T$ is the time of first entrance to $D^\comp = A\cup B$.
In this case, $-1/\log\lambda_1$ is on the order of $100$ for the lag times we consider and, as expected, inexact Richardson iteration converges slowly (Figure~\ref{fig:SI_Intro}, bottom left).  
Estimates of the committor by inexact Richardson iteration do not reach the correct values even after hundreds of iterations (Figure~\ref{fig:SI_Intro}, bottom right).

We now show that convergence can be accelerated dramatically by incorporating additional eigenfunctions of $\mathcal{S}^\tau$ (i.e., $k > 1$ in Algorithm \ref{alg:forecast}).
For the M\"uller-Brown model with a deepened intermediate basin, the second eigenvalue of $\mathcal{S}^\tau$ is of order $10^{-4}$ for a lag time of $\tau=1000$ steps or $1\,{\delta_t}^{-1}$  (while the first is near one as discussed above). We therefore choose $k=2$ with $\tilde{\varphi}^2_1$ initialized as a random linear combination of coordinate functions as in previous examples. 
We run the subspace iteration for four iterations, compute the committor as a linear combination of the resulting functions, and then refine this result with a further ten Richardson iterations (i.e., $k=1$ with the starting vector as the output of the $k=2$ subspace iteration).
To combine the functions, we use a linear solve which incorporates memory (Algorithm \ref{alg:memory}) \cite{darve_computing_2009,cao_advantages_2020}.  We find that the use of memory improves the data-efficiency substantially for poorly conditioned problems.  For our tests here, we use three memory kernels, corresponding to $\tau_{\rm mem}=\lfloor\tau / 4\rfloor$. 

\begin{figure*}
\begin{minipage}{\linewidth}
\begin{algorithm}[H]
\caption{Memory-corrected linear solve for predictions}
\label{alg:memory}
\begin{algorithmic}[1]
\Require{Stopped transition data $\set{X^0_j,X^{1 \bmin T_j}_j, \dots, X^{\tau \bmin T_j}_j}_{j=1}^n$, guess function $h$, reward data ${\set{ \rho(X_j)}}_{j=1}^n$, basis set $\{f^a\}_{a=1}^k$, lag between memory kernels $\tau_{\rm mem}$.}

\For{$s =0\dots (\tau / \tau_{\rm mem})$}
    \State Initialize the matrix $C^s$ with zeros \Comment{$C^s\in \mathbb{R}^{(k+1)\times (k+1)}$}
    \State $C_{11}^s\gets 1$
    \For{$a = 2\dots k$}
    \State $C_{a1}^s\gets \frac{1}{n}\sum_{j=1}^n f^a(X_j^0)\rho(X_j^{s\tau_{\rm mem} \bmin T_j})$
        \For{$b = 2\dots k$}
            \State $C_{ab}^s\gets \frac{1}{n}\sum_{j=1}^n f^a(X_j^0)f^b(X_j^{s\tau_{\rm mem} \bmin T_j})$
        \EndFor
    \EndFor
\EndFor
\State $A\gets C^1-C^0$
\For{$s = 0\dots (\tau / \tau_{\rm mem}) - 2$}
    \State $M^s\gets C^{s+2}-C^{s+1}-A(C^0)^{-1}C^{s+1}-\sum_{j=0}^{s}M^j(C^0)^{-1}C^{s-j}$
\EndFor
\State $A_{\text{mem}}\gets A+\sum_{s=0}^{(\tau / \tau_{\rm mem}) - 2} M^s$
\State Solve $A_{\text{mem}}w=w$
\State \Return $u=h +\sum_{a=2}^k w_a f^a$ 
\end{algorithmic}
\end{algorithm}
\end{minipage}
\end{figure*}

\begin{figure*}[hbtp]
    \centering
    \includegraphics[width=0.8\textwidth]{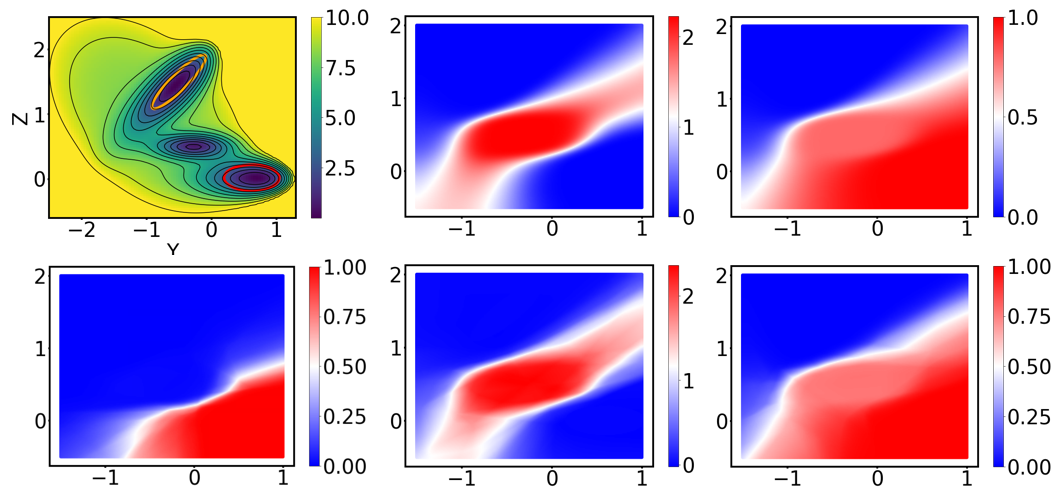}
    \caption{Illustration of the subspace iteration for the M\"uller-Brown committor.  (top left) Modified M\"uller-Brown potential. (top center) Reference second eigenfunction. (top right) Reference committor. (bottom left) Neural-network Richardson iterate after four iterations.  (bottom center) First non-dominant eigenfunction obtained from the neural network after four iterations. (bottom right) Committor resulting from linear combination of the Richardson iterate and second eigenfunction. Results shown are for $\tau=1000$ steps (i.e., $1\, {\delta_t}^{-1}$). 
    }
    \label{fig:SI_Illustration}
\end{figure*}

\begin{figure*}[bt]
    \centering
    \includegraphics[width=0.8\textwidth]{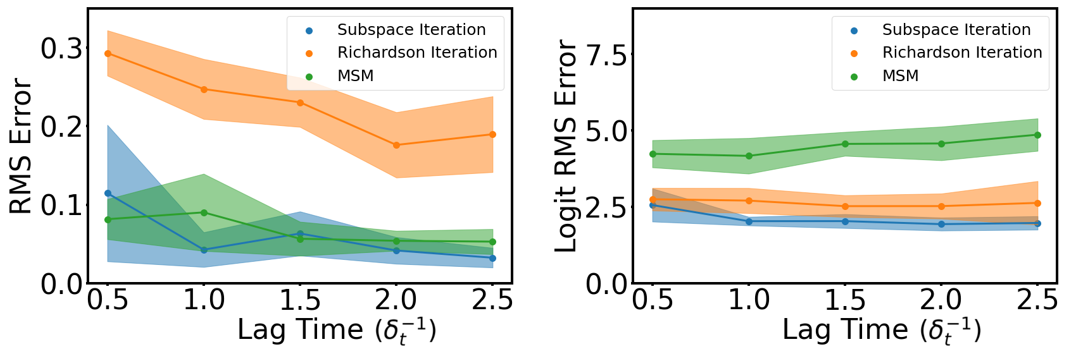}
    \caption{Committor for the M\"uller-Brown potential with deepened intermediate as a function of lag time (in units of ${\delta_t}^{-1}$). (left) Comparison of RMSE for subspace iteration as described above, Richardson iteration (as in  Section~\ref{sec:mbcommittor} but instead with 500 subspace iterations), and an MSM with 400 states.
    (right) RMSE of the logit function in \eqref{eq:logit}.}
    \label{fig:SI_Q_error}
\end{figure*}

The bottom row of Figure~\ref{fig:SI_Illustration} illustrates the idea of the subspace iteration.
The second eigenfunction (Figure \ref{fig:SI_Illustration}, center) is peaked at the intermediate.  As a result, the two neural-network functions linearly combined by the Galerkin approach with memory can yield a good result for the committor (Figure \ref{fig:SI_Illustration}, bottom right).
Figure~\ref{fig:SI_Q_error} compares the RMSE for the committor and the RMSE for the logit in \eqref{eq:logit} for Algorithm \ref{alg:forecast} with $k=1$ (pure Richardson iteration) and $k=2$ (incorporating the first non-trivial eigenfunction), and an MSM with 400 states.  We see that the Richardson iteration suffers large errors at all lag times; as noted previously, this error is mainly in the vicinity of the intermediate.  The MSM cannot accurately compute the small probabilities, but does as well as the subspace iteration in terms of RMSE.


\subsection{\texorpdfstring{AIB\textsubscript{9}}{AIB9} prediction results}\label{sec:aib9_results}

As an example of prediction in a high-dimensional system, we compute the committor for the transition between the left- and right-handed helices of AIB\textsubscript{9} using the inexact Richardson iteration scheme ($k=1$ in Algorithm~\ref{alg:forecast}) with the softplus loss function. Specifically, for this committor calculation $T$ is the time of first entrance to $D^\comp = A\cup B$ with $A$ and $B$ defined in Section~\ref{sec:aib_numerics}.
As before, we initialize $\tilde{\varphi}_1^1 = \ind{B}$.

To validate our results, we use the 5 $\mu$s reference trajectories to compute an empirical committor as a function of the neural network outputs, binned into intervals:
\begin{equation}
\label{eq:emp_q}
    \bar{q}(s) = \mathbbm{P}\left[X^T\in B\ \middle|\ u_\theta(X^0) \in [s, s + \Delta s]\right]
\end{equation}
for $s\in [0, 1 - \Delta s]$.
Here, we use $\Delta s = 0.05$.  
The
overall error in the committor estimate is defined as
\clearpage
\begin{equation}
\label{eq:emp_error}
q\ \textrm{error}=
    \left(\Delta s\sum_{n = 0}^{1 / \Delta s - 1} \big[\bar{q}(n\Delta s) - n\Delta s \big]^2 \right)^{1/2}.
\end{equation}
While this measure of error can only be used when the data set contains trajectories of long enough duration to reach $D^\comp$, it has the advantage that it does not depend on the choice of projection that we use to visualize the results.

\begin{figure*}[htb]
    \centering
    \includegraphics[width=0.8\textwidth]{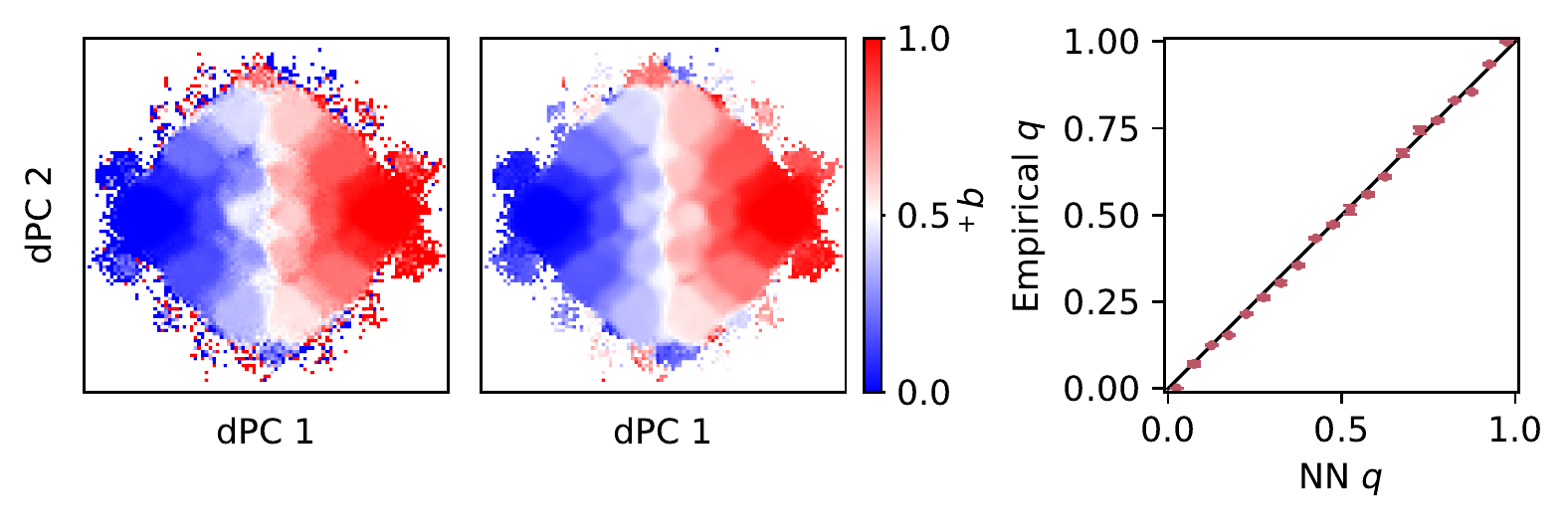}
    \caption{AIB\textsubscript{9} committor for the transition between left- and right-handed helices. 
    (left) Averages of $\mathbbm{1}_B(X^T)$ for initial conditions in bins in the first two dPCs computed from 20 long (5 $\mu$s) trajectories. 
    (middle) Averages of representative neural-network committors trained on the data set of 6,910 short (20 ns) trajectories; $\tau$ corresponds to 400 ps.
    (right) Comparison between empirical committors (as defined in \eqref{eq:emp_q}) and the neural-network committors (trained as for the middle panel). Error bars indicate standard deviations over ten different initializations of the neural-network parameters.}
    \label{fig:aib_q}
\end{figure*}

\begin{figure}[htb]
    \centering\includegraphics[width=\linewidth]{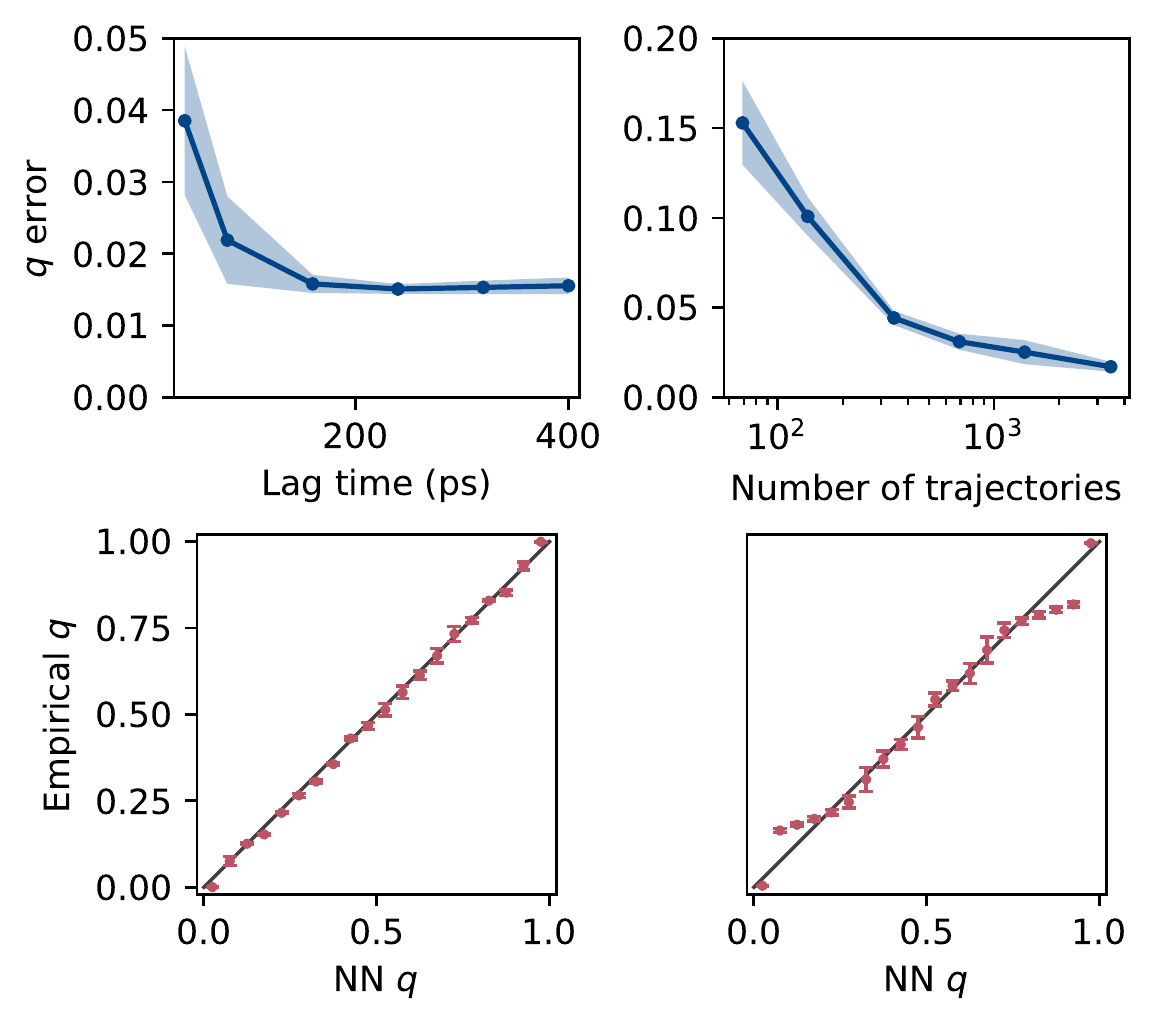}
    \caption{AIB\textsubscript{9} committor for the transition between left- and right-handed helices, as functions of lag time (in ps) and number of initial conditions. (top left) Error in the committor as a function of lag time  (in ps). Shading indicates the standard deviation over ten different initializations of the neural-network parameters.
    (top right) Error in the committor as a function of the number of initial conditions with $\tau$ corresponding to 160 ps.
    Shading indicates the standard deviation over ten different random samples of the trajectories.
    (bottom) Comparison between empirical committors and neural-network committors trained on data sets with (left) 1/2 and (right) 1/20 of the short trajectories. Error bars indicate standard deviations over ten random samples of the trajectories.}
    \label{fig:emp_error}
\end{figure}

Results for the full data set with $\tau$ corresponding to 400 ps are shown in Figure~\ref{fig:aib_q}.   The projection on the principal components is consistent with the symmetry of the system, and the predictions show good agreement with the empirical committors.  As $\tau$ decreases, the results become less accurate (Figure~\ref{fig:emp_error}, top left); at shorter lag times we would expect further increases in the error.
{ We also examine the dependence of the results on the size of the data set by subsampling the short trajectories and then training neural networks on the reduced set of trajectories (Figure~\ref{fig:emp_error}, top right).
We find that the performance steadily drops as the number of trajectories is reduced and degrades rapidly for the data sets subsampled more than 20-fold (Figure~\ref{fig:emp_error}, bottom), corresponding to about 7 $\mu$s of total sampling.}

Finally, we compute the MFPT to reach the right-handed helix using the same data set.
For the MFPT calculation $T$ is the time of first entrance to $D^\comp=B$.  Note that the time of first entrance to $B$ includes long dwell times in $A$ and is expected to be much larger than the time of first entrance to $A\cup B$.
 
 We compare against an empirical estimate of the MFPT defined by
\begin{equation}
\label{eq:emp_mfpt}
    \bar{m}(s) = \E\left[T \middle| u_\theta(X^0) \in [s, s + \Delta s]\right]
\end{equation}
for $s\in [0, m_{\max} - \Delta s]$ where $\Delta s=3$ and $m_{\max} = $ 57~ns.
Overall error is defined analogously to Eq.~\eqref{eq:emp_error}.

\begin{figure*}[hbt]
    \centering
    \includegraphics[width=0.8\textwidth] {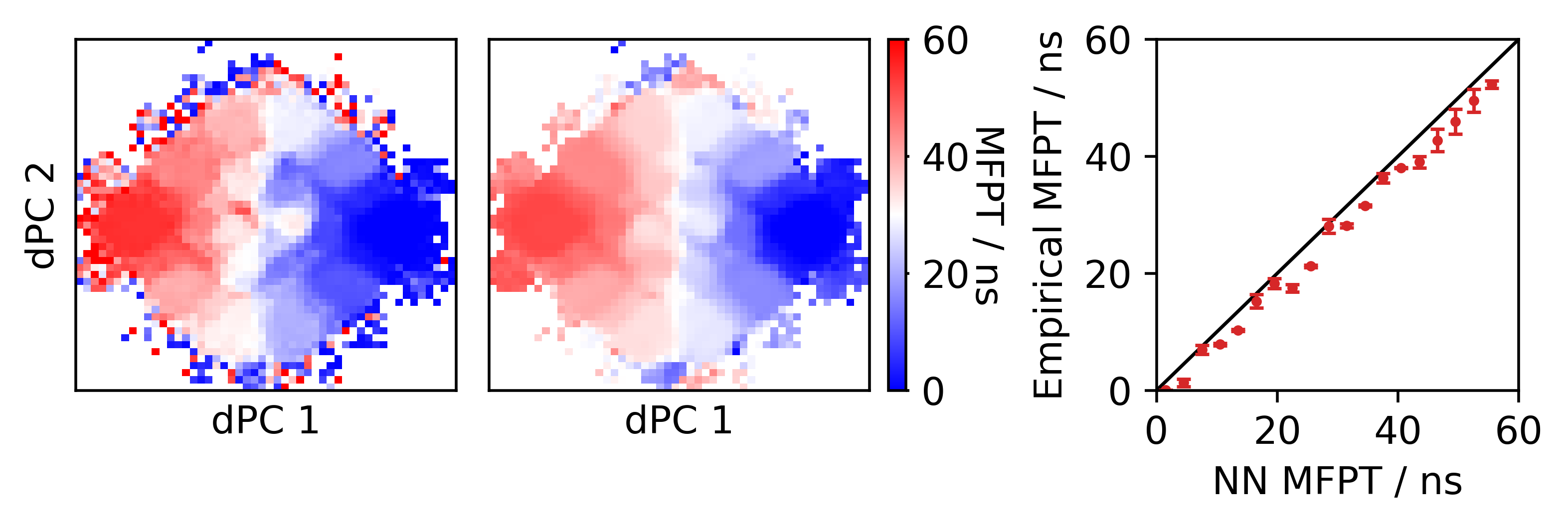}
    \caption{AIB\textsubscript{9} MFPT to the right-handed helix. 
    (left) Averages of the time to next reach $B$ for initial conditions in bins in the first two dPCs computed from 20 long (5 $\mu$s) trajectories. 
    (middle) Averages of representative neural-network committors trained on the data set of 6,910 short (20 ns) trajectories; $\tau$ corresponds to 400 ps.
    (right) Comparison between empirical committors (as defined in \eqref{eq:emp_mfpt}) and the neural-network committors (trained as for the middle panel).
    Error bars indicate standard deviations over ten different initializations of the neural-network parameters.}
    \label{fig:aib_mfpt}
\end{figure*}

\begin{figure*}[hbt]
    \centering
    \includegraphics[width=0.8\textwidth] {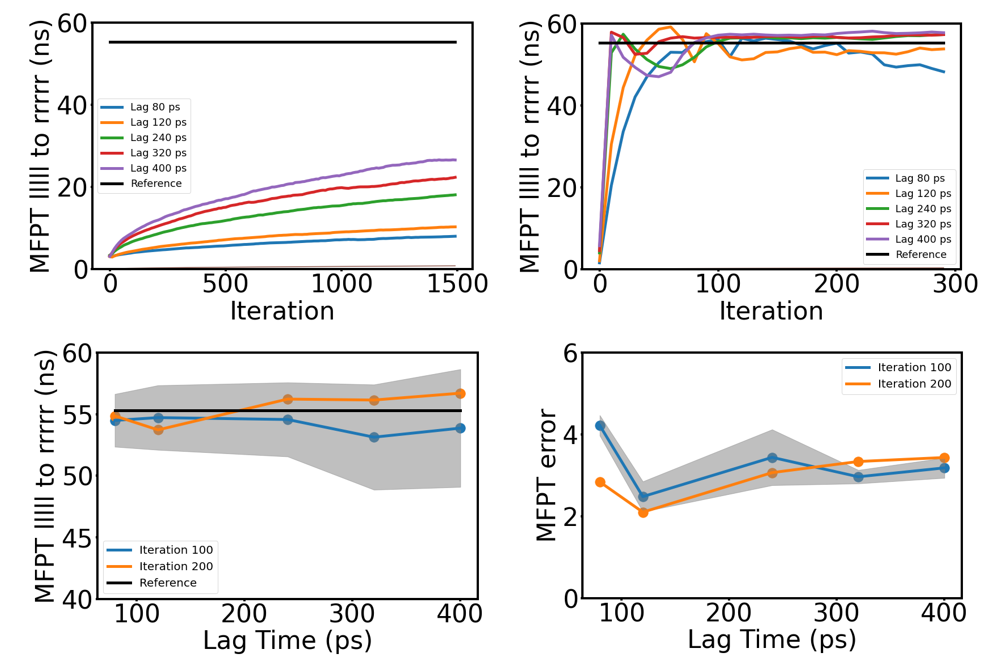}
    \caption{
    MFPT between left- and right-handed helices for the AIB\textsubscript{9} system. (top left) Convergence of Richardson iteration.  The lllll to rrrrr MFPT is computed by averaging the richardson iteration result over each frame of each of the long reference trajectories in the lllll state. (top right) Convergence of a five-dimensional subspace iteration.  (bottom left) MFPT after 100 and 200 subspace iterations as a function of lag time.  Shading indicates standard deviations over ten different initializations of the neural-network parameters. (bottom right) Overall error in MFPT.
    To obtain the results shown in this figure, we first use the short-trajectory dataset to train neural networks to predict the MFPT; we then use these networks with fixed parameters to evaluate the MFPT for all structures in the long reference trajectories and average the results for those structures in the left-handed helix state.  The horizontal lines in the top panels are obtained from averaging the time to the right-handed helix for the same structures.
    }
    \label{fig:aib9_mfpt}
\end{figure*}

In Figure~\ref{fig:aib_mfpt}, we show the MFPT obtained from Algorithm \ref{alg:forecast} with $k=5$ and the $L^2_\mu$ loss function. 
Initially we set $\tilde{\varphi}^1_1$ equal to an arbitrary positive function (we use $5\mathbbm{1}_A$) and $\tilde{\varphi}^a_s$ for $a>1$ to a random linear combination of coordinate functions. In Figure~\ref{fig:aib9_mfpt} we examine the convergence of the MFPT from the left-handed helix to the right-handed helix for the MFPT computed with $k=1$ (pure Richardson iteration)
and $k=5$.
The horizontal line indicates a MFPT of about 56~ns estimated from the long reference trajectories.
We see that the algorithm with $k=5$ converges much faster (note the differing scales of the horizontal axes) and yields accurate results at all lag times other than the shortest shown. The need to choose $k>1$ for this MFPT calculation is again consistent with theoretical convergence behavior of exact subspace iteration.  Because the typical time of first entrance to $B$ from points in $A$ is very large, we expect the dominant eigenvalue of $\mathcal{S}^\tau$ to be very near to one when $D=B^\comp$.  In contrast, the committor calculation benefits from the fact that the time of first entrance to $A\cup B$ is much shorter, implying a smaller dominant eigenvalue of $\mathcal{S}^\tau$ when $D=(A\cup B)^\comp.$


\section{Conclusions}

In this work we have presented a method for spectral estimation and rare-event prediction from short-trajectory data.  The key idea is that we use the data as the basis for an inexact subspace iteration.  For the test systems that we considered, the method not only outperforms high-resolution MSMs, but it can be tuned through the choice of loss function to compute committor probabilities accurately near the reactants, transition states, and products.  Other than the Markov assumption, our method requires no knowledge of the underlying model and puts no restrictions on its dynamics. 

As discussed in prior neural-network based prediction work \cite{strahan2023predicting, rotskoff_active_2022}, our method is sensitive to the quality and distribution of the initial sampling data. However, our work shares with Ref.~ \citenum{strahan2023predicting} the major advantage of allowing the use of arbitrary inner products.  This enables adaptive sampling of the state space \cite{lucente_coupling_2022, strahan2023predicting} and---together with the features described above---the application to observational and experimental data, for which the stationary distribution is generally unavailable.



In the present work, we focused on statistics of transition operators, but our method can readily be extended to solve problems involving their adjoint operators as well.
By this means, we can obtain the stationary distribution as well as the backward committor.  The combination of forward and backward predictions allows the analysis of transition paths using transition path theory without needing to generate full transition paths \cite{e_transition-path_2010, vanden2006transition,lorpaiboon_augmented_2022} and has been used to understand rare transition events in molecular dynamics \cite{strahan_long-time-scale_2021, noe_constructing_2009, antoszewski_kinetics_2021, guo_dynamics_2022, meng_transition_2016, vani2022computing} and geophysical flows \cite{Finkel2020paths, miron_transition_2021, lucente_committor_2022, finkel2023revealing, finkel_data-driven_2023}.
We leave these extensions to future work.


In cases in which trajectories that reach the reactant and product states are available, it would be interesting to compare our inexact iterative schemes against schemes for committor approximation based on logistic regression and related approaches \cite{Ma2005nn, peters_obtaining_2006, peters2007extensions, hu2008two, jung_artificial_2019, chattopadhyay2020analog, jung_machine-guided_2023, miloshevich_probabilistic_2023}.  These schemes are closely related to what is called ``Monte-Carlo'' approximation in reinforcement learning \cite{sutton_reinforcement_2018}, and also to the inexact Richardson iteration that we propose here with $\tau \rightarrow \infty$.

We have seen that temporal difference (TD) learning, a workhorse for prediction in reinforcement learning, is closely related to an inexact form of Richardson iteration. 
Variants like TD$(\lambda)$, have similar relationships to inexact iterative schemes.   As we showed, subspace iteration is a natural way of addressing slow convergence.  We thus anticipate that our results have implications for reinforcement learning, particularly in scenarios in which the value function depends on the occurrence of some rare-event. 
Finally, we note that our framework can be extended to the wider range of iterative numerical linear algebra algorithms.   In particular, Krylov or block Krylov subspace methods may offer further acceleration.  In fact, very recently an approach along these lines was introduced for value function estimation in reinforcement learning \cite{pmlr-v206-xia23a}.


%

\begin{acknowledgements}
We are grateful to Arnaud Doucet for pointing our attention to TD methods and the inexact power iteration in Ref.\ \onlinecite{wen_batch_2020}, both of which were key motivations for this work. We also thank Joan Bruna for helpful conversations about reinforcement learning.
This work was supported by National Institutes of Health award R35 GM136381 and National Science Foundation award DMS-2054306.
S.C.G.\ acknowledges support by the National Science Foundation Graduate Research Fellowship under Grant No.\ 2140001. J.W.\  acknowledges support from the Advanced Scientific Computing Research Program within the DOE Office of Science through award DE-SC0020427.
This work was completed in part with resources provided by the University of Chicago Research Computing Center, and we are grateful for their assistance with the calculations.
 ``Beagle-3: A Shared GPU Cluster for Biomolecular Sciences'' is supported by the National Institute of Health (NIH) under the High-End Instrumentation (HEI) grant program award 1S10OD028655-0.
\end{acknowledgements}

\section*{Data Availability Statement}
The data that supports the findings of this study are available within the article. Code implementing the algorithms and a Jupyter notebook illustrating use of the method on the M\"uller-Brown example are available at https://github.com/dinner-group/inexact-subspace-iteration.

\section*{References}

\bibliographystyle{unsrt}
\bibliography{ref}


\end{document}